\documentclass[useAMS,usenatbib]{mn2e}

\usepackage[T1]{fontenc}
\usepackage{aecompl}
\usepackage{subcaption}
\usepackage{amsmath}
\usepackage{amssymb}
\usepackage{amsmath}
\usepackage{graphicx}
\usepackage{url}
\newcommand{\Msol}{M_{\odot}}
\newcommand{\pl}{\texttt{PixeLens}}
\newcommand{\gl}{\texttt{glafic}}

\title[Strong lensing in RX J1347.5-1145 revisited]{Strong lensing in RX J1347.5-1145 revisited\thanks{Based on observations made with the NASA/ESA Hubble Space Telescope, which is operated by the Association of Universities for Research in Astronomy, Inc., under NASA contract
NAS 5-26555.}}
\author[F. K\"ohlinger and R. W. Schmidt]{F. K\"ohlinger$^{1,2}$\thanks{E-mail:
fkoehlin@strw.leidenuniv.nl} and R. W.
Schmidt$^{1}$\\
$^{1}$Astronomisches Rechen-Institut, Zentrum f\"ur Astronomie der Universit\"at Heidelberg, M\"onchhofstr. 12-14, Heidelberg, 69120, Germany\\
$^{2}$Leiden Observatory, Leiden University, Niels Bohrweg 2, Leiden, NL-2333 CA, The Netherlands}

\begin{document}

\date{Accepted 0000 XXXX 00. Received 0000 XXXX 00; in original form 2013 XXXX 00}

\pagerange{\pageref{firstpage}--\pageref{lastpage}} \pubyear{2013}

\maketitle

\label{firstpage}

\begin{abstract}

We present a revised strong lensing mass reconstruction of the galaxy cluster RX J1347.5-1145. The X-ray luminous cluster at redshift $z=0.451$ has already been studied intensively in the past. Based on information of two such previous (strong-)lensing studies by \citet{Halkola08} and \citet{Bradac08}, as well as by incorporating newly available data from the \textit{Cluster Lensing And Supernovae survey with Hubble} (CLASH, \citealt{Postman12}), we identified four systems of multiply lensed images (anew) in the redshift range $1.75 \leq z \leq 4.19$. One multiple image system consists of in total eight multiply lensed images of the same source. The analysis based on a parametric mass model derived with the software \gl\ \citep{Oguri10} suggests that the high image multiplicity is due to the source ($z_{phot} = 4.19$) being located on a so-called ''swallowtail`` caustic. In addition to the parametric mass model, we also employed a non-parametric approach using the software \pl\ \citep{Saha97, Saha04} in order to reconstruct the projected mass of the cluster using the same strong lensing data input. 

Both reconstructed mass models agree in revealing several mass components and a highly elliptic shape of the mass distribution. Furthermore, the projected mass inside, for example, a radius $R \sim 35'' \sim 200 \, \mathrm{kpc}$ of the cluster for a source at redshift $z=1.75$ is $M(<R) \approx (2.19^{+0.01}_{-0.02}) \times 10^{14} \, \mathrm{\Msol}$ as estimated by \gl . Within the same radius \pl\ predicts a mass of $M(<R) \approx (2.47 \pm 0.01) \times 10^{14} \, \mathrm{\Msol}$ which exceeds the \gl\ estimate by $\sim 13 \, \%$. The difference could be related to the fundamental degeneracy involved when constraining dark matter substructures with gravitationally lensed arcs.
\end{abstract}


\section{Introduction}
\label{sec:intro}
In resolving the nature of the two exotic ingredients of current standard $\Lambda$CDM cosmology - dark matter and dark energy - the determination of accurate masses and mass profiles plays an important role: the mass distribution in galaxy clusters, for example, is a direct test for predictions of the cold dark matter (CDM) paradigm (e.g. \citealt{Bartelmann13}) since numerical simulations in the scope of $\Lambda$CDM predict a universal mass profile for mass haloes covering scales from galaxies to clusters of galaxies (\citealt{Navarro97, Navarro10}).

Once the mass and its distribution are determined, dark energy models can be constrained with various techniques either from using mass calibrated number counts or scaling relations (e.g. \citealt{Allen11, Giodini13, Planck13} and references therein) or even using the systems on their own (e.g. \citealt{Golse02}, \citealt{Jullo10}).
Therefore, mass profiles of galaxy clusters are a valuable probe for putting further constraints on cosmological parameters such as $\Omega_m$, $\sigma_8$ or the equation of state parameter for dark energy $w$.

In this paper we scrutinize the evidence for multiple images of background systems and its implication for the central mass distribution in the system RX J1347.5-1145.

\subsection{RX J1347.5-1145}

The galaxy cluster RX J1347.5-1145 at redshift $z = 0.451$ is among the most luminous X-ray clusters known to date \citep{Schindler95} and has already been studied intensively in the past.

Various data sets are available for this cluster ranging from X-ray \citep{Schindler95, Schindler97, Allen02, Ettori04, GittiSchindler04, Gitti07, Mahdavi13} to optical \citep{FischerTyson97, Sahu98, CohenKneib02, RavindranathHo02, Verdugo12} and radio observations of the Sunyaev-Zel'dovich (SZ) effect \citep{Pointecouteau01, Komatsu01, Kitayama04, Plagge13}.
Mass estimates from various studies using different techniques like strong-lensing, weak-lensing, a combined strong and weak lensing analysis, X-ray measurements, velocity dispersion measurements from spectroscopic data or the SZ effect often yielded discrepant results. 
Particularly, the dynamical mass estimate \citep{CohenKneib02}, early X-ray measurements \citep{Schindler97} and gravitational lensing estimates \citep{FischerTyson97} yielded a discrepancy of factor $\sim 3$. 
  
Possible reasons for the discrepancy between the results of the different mass reconstruction approaches might be due to the shape of the cluster potential, projection effects or more complicated gas physics in the cluster which have not been fully taken into account. Moreover, the different mass reconstruction techniques are also affected by various systematic effects such as cluster member contamination, unknown or uncertain redshifts for multiply lensed images, ambiguous identification of multiple image systems and temperature calibrations, and it is very challenging to quantify the effects of these systematics correctly on the uncertainties of the respective analysis.

A specific proposal for resolving the discrepancy between the mass estimates in RX J1347.5-1145 was suggested by \citet{CohenKneib02} stating that the cluster is likely to be ongoing a major merger. This would bias the velocity dispersion measurements and also affect the X-ray results due to special merger dynamics and thus might be the key to reconciling the low velocity dispersion mass estimate with the higher estimates from lensing and X-ray measurements. 

\begin{figure*}
    \centering
    \begin{minipage}{140mm}
        \centering
        \includegraphics[width=\textwidth]{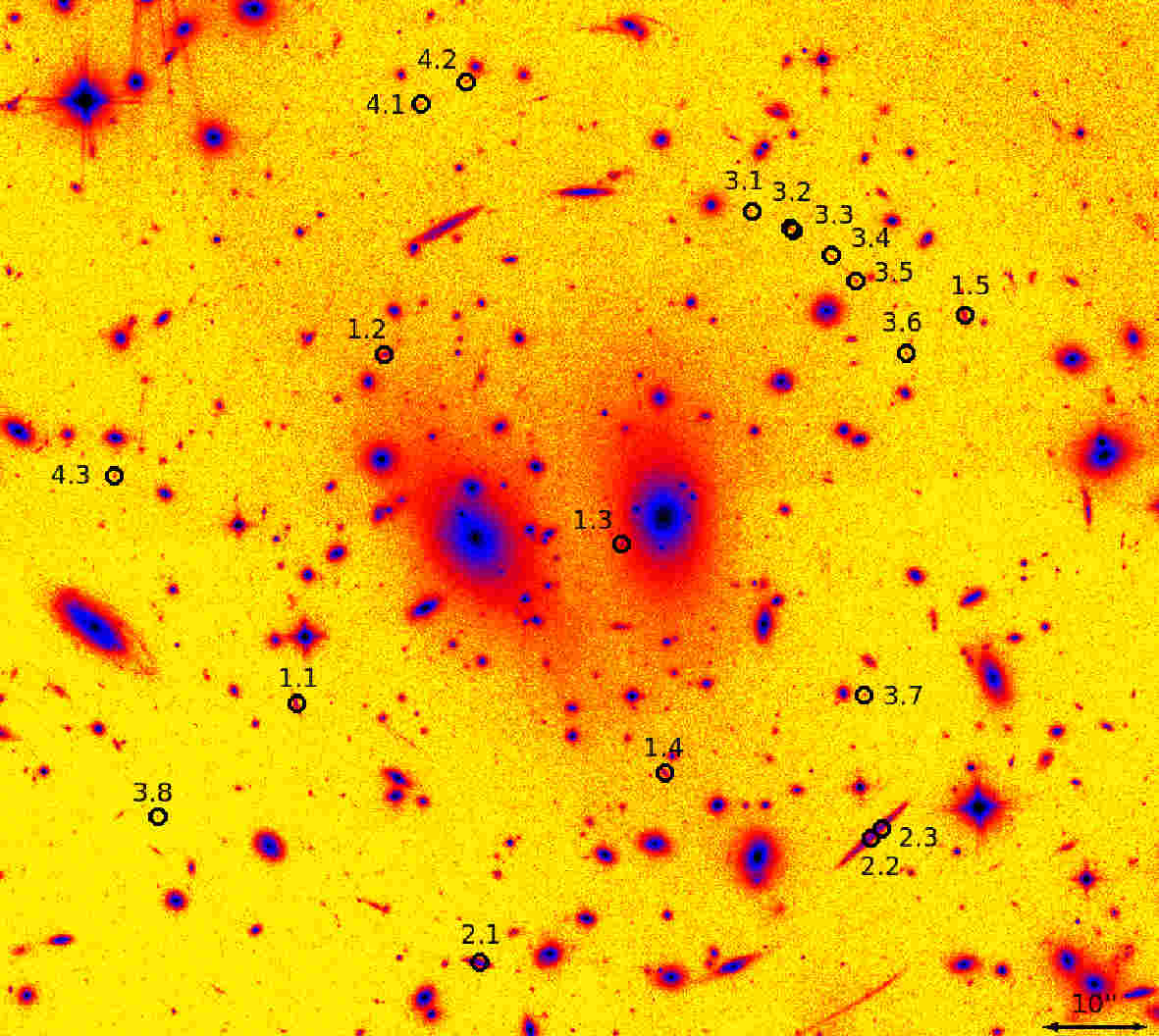}
        \caption{The CLASH ACS-IR detection image (cf. Sect. 
                 \ref{sec:data}). North is up and East is left. For 
                 better visibility of faint multiply lensed images, this 
                 image is presented in false colours which encode 
                 brightness information. Denoted are identified         
                 multiple image system candidates as used in this work 
                 for the best-fitting models discussed in Sect. 
                 \ref{sec:analysis}. 
                 Also compare with identifications of multiple image 
                 systems from \citet{Bradac08} and \citet{Halkola08} in 
                 Tab.~\ref{tab:arcs}.}
        \label{fig:all_arcs}
    \end{minipage}
\end{figure*}

The peculiarity of the cluster RX J1347.5-1145 to contain \textit{two} bright cD galaxies (cf. Fig.~\ref{fig:all_arcs}; throughout this work we will refer to the western cD galaxy as BCG while calling the other just second cD galaxy) at its centre can be taken as further support for this merger scenario. Moreover, a region of shocked gas in the south-east of the cluster was observed by \citet{Komatsu01} using the SZ effect and by \citet{Allen02} in X-ray data from \textit{Chandra} (cf. Fig.~\ref{fig:kappa_gl}).

All these peculiarities and especially the discrepancies in the mass reconstruction of those early papers were revised when new deep data from the \textit{Advanced Camera for Surveys} (ACS) from the \textit{Hubble Space Telescope} (HST) became available in 2008. \citet{Bradac08} presented a mass reconstruction based on a combined strong and weak lensing analysis and found their results in accordance to results obtained from an X-ray analysis, also presented in their paper (cf. Fig.~\ref{fig:mass_all}). \citet{Halkola08} obtained results solely based on a strong lensing analysis, but also found their mass results to agree with other results presented in the literature (cf. Fig.~\ref{fig:mass_all}).

Despite the consistent mass estimates, both studies did not quite present a consistent description of the cluster with respect to its strong lensing features. First of all, there is a somewhat ambiguous and sometimes even contradictory identification of multiple image systems between both papers (cf. Tab.~\ref{tab:arcs}).
Moreover, in some of the identified multiple image systems, there are some missing but necessarily expected counter images which will be discussed in more detail in Sect. \ref{sec:mult_img}.

Another point was, that there was only one spectroscopically confirmed redshift for only one multiple image system available at the time of their analyses. The redshifts of other multiple image systems were either fixed using lower precision photometric redshifts or left free for fitting then. Therefore, as soon as high quality data from CLASH \citep{Postman12} became available, we started to investigate the cluster again with strong lensing in order to present a more consistent high detail strong lensing model. As mentioned before crucial ingredients for a strong lensing mass reconstruction are fixed redshifts of a multiple image system and the unambiguous identification of such systems. With the high quality of the provided CLASH data including robust photometry and reliable photometric redshifts and by having improved the multiple image identification with this data, we present here a high detail strong lensing model of the cluster RX J1347.5-1145.
This entails a thorough discussion of the known multiply imaged systems, but fortunately does not alter the consistency with X-ray mass models when allowing for substructure in the core.
  
The paper is structured as follows. In Section \ref{sec:data} we describe the input data used for our analysis and explain the data fitting further in Section \ref{sec:methods}. The analysis and results are presented in Section \ref{sec:analysis} and conclusions are finally drawn in Section \ref{sec:conclusion}.  
   
For comparison of results obtained here with previous work we also adopt a $\Lambda$CDM cosmology with $\Omega_m = 0.3$, $\Omega_{\Lambda} = 0.7$ and Hubble constant $\text{H}_0=70 \, \mathrm{km} \, \mathrm{s}^{-1} \, \mathrm{Mpc}^{-1}$. At the redshift of the cluster, $z_{lens} = 0.451$, one arcsecond then corresponds to $5.77 \, \mathrm{kpc}$.

\section{Data}
\label{sec:data}

The galaxy cluster RX J1347.5-1145 was reobserved as one out of $25$ galaxy cluster targets of the CLASH programme (cf. \citealt{Postman12} for full details on this survey).

The images were taken in $16$ broadband filters of the \textit{Wide Field Camera 3} (WFC3)/UVIS\footnote{F$225$W, F$275$W, F$336$W and F$390$W}, WFC3/IR\footnote{F$105$W, F$110$W, F$125$W, F$140$W and F$160$W} and ACS/WFC\footnote{F$435$W, F$475$W, F$606$W, F$625$W, F$775$W, F$814$W and F$850$LP} of HST comprising a total spectral range from near-UV to near-IR, yielding highly reliable photometry. The filters were especially selected for that purpose based on tests with simulated photometric data in order to achieve a precision on photometric redshifts of $\sigma_z \sim 0.02(1+z)$ using the \textit{Bayesian photometric redshift} (BPZ) package \citep{Benitez00, Benitez04, Coe06}.

In addition to images, catalogues including the photometry from all $16$ bands and photometric redshifts for identified sources are also made available to the public.   
For the creation of these public catalogues one detection image consisting of the weighted sum of all ACS/WFC and WFC3/IR images was used in order to run the software \textit{SExtractor} \citep{Bertin96} on it for detecting objects and measuring their photometry. A second detection image was created solely from the WFC3/IR images for the search for highly redshifted objects, also resulting in a respective second catalogue. Both catalogues are based on the lower-resolution (i.e. $1 \, \mathrm{pix} = 65 \, \mathrm{mas}$) detection images. We will refer to the first catalogue based on ACS/WFC and WFC3/IR images as ''CLASH ACS-IR catalogue`` and to the second one based on WFC3/IR images only as ''CLASH IR catalogue``. The detection images are denoted as ''CLASH ACS-IR detection image`` and ''CLASH IR detection image``, respectively.

In the following analysis we primarily consulted the CLASH ACS-IR catalogue due to the more rigid criteria on object detections. Just for objects not already contained in there (chiefly multiple image system 1), we used the CLASH IR catalogue. 

\subsection{Multiple image systems in RX J1347.5-1145} 
\label{sec:mult_img}

As mentioned in Sect. \ref{sec:intro}, there are already candidates for multiple image systems published in the literature \citep{Bradac08, Halkola08}. These served as initial data input for our early models and we usually followed the identifications and image affiliations by \citet{Halkola08}.

However, the CLASH data and catalogues and also predictions from early models led to a new interpretation of some of the multiple image systems. Moreover, we could also include additional images that were not yet included in previous studies.

We do not consider single image systems of previous studies, because these do not impose any new constraints on the strong lensing model. We show an overview of all images used during our final analysis in Fig.~\ref{fig:all_arcs}. In Tab.~\ref{tab:arcs} we summarize the properties of these systems and also indicate for ease of comparison their nomenclature in the studies of \citet{Halkola08} and \citet{Bradac08}, if applicable. 

The cluster centre is set on the BCG at position $\text{RA}=206.8775 \, \deg$, $\text{Dec}=-11.7526 \, \deg$ (J2000). Furthermore, all photometric redshift estimates for the multiple image systems are derived as an average over the individual redshifts (cf. Tab.~\ref{tab:arcs}) of the images affiliated to the respective multiple image system. Additionally, we checked colours and surface brightnesses based on the 16 band photometry of all multiple image systems to be consistent. Exemplarily, we list one colour ($M_{F475W}-M_{F814W}$) and the surface brightness in one filter ($S_{F814W}$) in Tab.~\ref{tab:arcs} for each multiply lensed image.

\begin{table*}
\centering 
\begin{minipage}{175mm}
  \caption{Positions ($\Delta$RA, $\Delta$Dec), colour ($M_{F475W}-M_{F814W}$), surface brightness ($S_{F814W}$) and photometric redshifts     
           ($z_{phot}$) with $95 \, \%$ confidence intervals for each image of the multiple image systems from the indicated CLASH catalogues. 	             
           Note that image 1.3 is apparently a drop out (very closely located to BCG) and could not be identified in the CLASH ACS-IR 
           catalogue either. Therefore, we did not consider it for the calculation of the redshift of the whole system. All coordinates are given 
           relative to the cluster centre at position $\text{RA}=206.8775 \, \deg$, $\text{Dec}=-11.7526 \, \deg$ (J$2000$). Please refer also to 
           Fig.~\ref{fig:all_arcs}. For convenience we also give the nomenclature from \citet{Halkola08} [Ha08] and \citet{Bradac08} [Br08].}
  \label{tab:arcs}
  \tabcolsep=0.11cm
   \begin{tabular}{ccccccccccc}
\hline
\multicolumn{3}{c}{Multiple Images} & \multicolumn{ 1}{c}{$\Delta$RA $['']$} & \multicolumn{ 1}{c}{$\Delta$Dec $['']$} & \multicolumn{ 1}{c}{$M_{F475W}-M_{F814W}$} & \multicolumn{ 1}{c}{$S_{F814W} \, [mag/arcsec^2]$} & \multicolumn{ 1}{c}{$z_{phot}$} & \multicolumn{ 1}{c}{$z_{spec}$} & \multicolumn{ 1}{c}{catalogue} & \multicolumn{ 1}{c}{comments} \\ \cline{ 1- 3}

This work & Ha08 & Br08 & \multicolumn{ 1}{l}{} & \multicolumn{ 1}{l}{} & \multicolumn{ 1}{l}{} & \multicolumn{ 1}{l}{} & \multicolumn{ 1}{l}{} & \multicolumn{ 1}{l}{} & \multicolumn{ 1}{l}{} \\ \hline

\parbox[0pt][1.5em][c]{0cm}{}  1.1 &  1a & I & $-35.61$ & $-17.70$ & $0.20 \pm 0.05$ & $24.42 \pm 0.03$ & $2.39^{+0.03}_{-0.08}$ & - & IR & \ref{foot:phot}\\
\parbox[0pt][1.5em][c]{0cm}{}  1.2 &  1b & I & $-26.96$ & $15.34$ & $0.06 \pm 0.05$ & $24.55 \pm 0.03$ & $2.19^{+0.01}_{-0.13}$ & - & IR & \ref{foot:phot}\\
\parbox[0pt][1.5em][c]{0cm}{}  1.3 &  1c & I & $-4.02$ & $-2.54$ & $1.53 \pm 0.17$ & $24.71 \pm 0.03$ & $0.55^{+0.01}_{-0.06}$ & - & IR & \footnote{\label{foot:phot}The error bars of the faint IR catalogue magnitudes appear somewhat underestimated.}, \footnote{\label{foot:1.3}Affected by BCG}\\
\parbox[0pt][1.5em][c]{0cm}{}  1.4 &  1d & I & $0.01$ & $-24.24$ & $0.29 \pm 0.06$ & $24.35 \pm 0.03$ & $1.92^{+0.14}_{-0.04}$ & - & IR &\ref{foot:phot}\\
\parbox[0pt][1.5em][c]{0cm}{}  1.5 &  1e & I & $28.97$ & $19.06$ & $0.15 \pm 0.06$ & $24.54 \pm 0.03$ & $2.39^{+0.03}_{-0.24}$ & - & IR &\ref{foot:phot}\\ \hline
\parbox[0pt][1.5em][c]{0cm}{}  2.1 &  2a & A & $-18.00$ & $-42.19$ & $0.20 \pm 0.02$ & $23.48 \pm 0.01$ & $1.78^{+0.01}_{-0.01}$ & $1.75$ &ACS-IR & \\
\parbox[0pt][1.5em][c]{0cm}{}  2.2 &   - & - & $19.55$ & $-30.46$ & - & - & - & - & - & \footnote{\label{foot:2.2}The object appears as a single source in the catalogue at position 2.3. Due to basic lensing geometry, we decided to split the object up into two components (cf. Sect.~\ref{sec:img_sys2}).}\\
\parbox[0pt][1.5em][c]{0cm}{}  2.3 &  2b & A & $20.57$ & $-29.46$ & $0.18 \pm 0.01$ & $23.47 \pm 0.01$ & $1.78^{+0.01}_{-0.01}$ & $1.75$ &ACS-IR & \ref{foot:2.2}
\\ \hline
\parbox[0pt][1.5em][c]{0cm}{}  3.1 & 12a & B & $8.27$ & $28.92$ & $1.62 \pm 0.40$ & $23.57 \pm 0.09$ & $4.01^{+0.18}_{-0.19}$ & - & ACS-IR & \\
\parbox[0pt][1.5em][c]{0cm}{}  3.2 &  -  & - & $12.15$ & $27.32$ & $1.51 \pm 0.33$ & $23.75 \pm 0.08$ & $4.11^{+0.13}_{-0.17}$ & - &ACS-IR & \footnote{\label{foot:3.2}SExtractor may not split the object correctly.}\\
\parbox[0pt][1.5em][c]{0cm}{}  3.3 &  -  & - & $12.45$ & $27.14$ & $3.01 \pm 0.74$ & $23.41 \pm 0.06$ & $4.36^{+0.08}_{-0.10}$ & - &ACS-IR &\ref{foot:3.2} \\
\parbox[0pt][1.5em][c]{0cm}{}  3.4 & 12b & B & $16.13$ & $24.72$ & $2.23 \pm 0.47$ & $23.34 \pm 0.07$ & $4.14^{+0.11}_{-0.12}$ & - &ACS-IR & \\
\parbox[0pt][1.5em][c]{0cm}{}  3.5 & 11a & B & $18.51$ & $22.36$ & $2.00 \pm 0.34$ & $23.42 \pm 0.06$ & $4.23^{+0.11}_{-0.05}$ & - &ACS-IR & \\
\parbox[0pt][1.5em][c]{0cm}{}  3.6 & 11b & B & $23.35$ & $15.51$ & $1.76 \pm 0.26$ & $23.52 \pm 0.05$ & $4.15^{+0.08}_{-0.08}$ & - &ACS-IR & \\
\parbox[0pt][1.5em][c]{0cm}{}  3.7 & 11c?& - & $19.38$ & $-16.84$ & $2.06 \pm 0.39$ & $23.50 \pm 0.07$ & $4.39^{+0.10}_{-0.11}$ & - &ACS-IR & \\
\parbox[0pt][1.5em][c]{0cm}{}  3.8 & 11d?& - & $-48.95$ & $-28.36$ & $2.44 \pm 0.81$ & $23.49 \pm 0.11$ & $4.16^{+0.14}_{-0.31}$ & - &ACS-IR & \\ \hline
\parbox[0pt][1.5em][c]{0cm}{}  4.1 &  8a & C & $-23.56$ & $39.08$ & $0.98 \pm 0.17$ & $23.59 \pm 0.06$ & $3.57^{+0.14}_{-0.06}$ & - &ACS-IR & \\
\parbox[0pt][1.5em][c]{0cm}{}  4.2 &  8b & C & $-19.05$ & $41.27$ & $1.00 \pm 0.14$ & $23.87 \pm 0.05$ & $3.62^{+0.12}_{-0.06}$ & - &ACS-IR & \\
\parbox[0pt][1.5em][c]{0cm}{}  4.3 &  8c?& - & $-53.14$ & $3.95$ & $1.01 \pm 0.18$ & $23.76 \pm 0.07$ & $3.71^{+0.07}_{-0.12}$ & - &ACS-IR & \\  
\hline
\end{tabular}
\end{minipage}
\end{table*}

\subsubsection{Multiple image system 1 (5 images)}

This system consists of five images on four sides of the cluster and one central image (cf. Fig.~\ref{fig:all_arcs}). It was already identified in \citet{Bradac08} and \citet{Halkola08}, and the photometric redshift of this system estimated from the CLASH IR catalogue yields $z_{phot}=2.22^{+0.05}_{-0.12}$ which is consistent with the redshift estimates for images ''1d`` and ''1e`` by \citet{Halkola08} with $z^{1d}_{phot} = 2.19 \pm 0.05$ and $z^{1e}_{phot} = 2.19 \pm 0.15$, respectively (cf. Tab.~\ref{tab:arcs} for nomenclature). The redshift estimate by \citet{Bradac08} from fitting strong-lensed data yields $z^{I}_{fit} = 1.7 \pm 0.2$ though (cf. Tab.~\ref{tab:arcs} for nomenclature).

Note that we did not include image 1.3 in the redshift average, though being a likely member of the multiple image system, the corresponding photometry seems to be affected by the BCG (cf. Fig \ref{fig:all_arcs}).

\subsubsection{Multiple image system 2 (3 images)}
\label{sec:img_sys2}

\begin{figure*}
    \begin{minipage}{140mm}
    \centering
    \begin{subfigure}{0.49\textwidth}
        \centering
        \includegraphics[width=\textwidth]{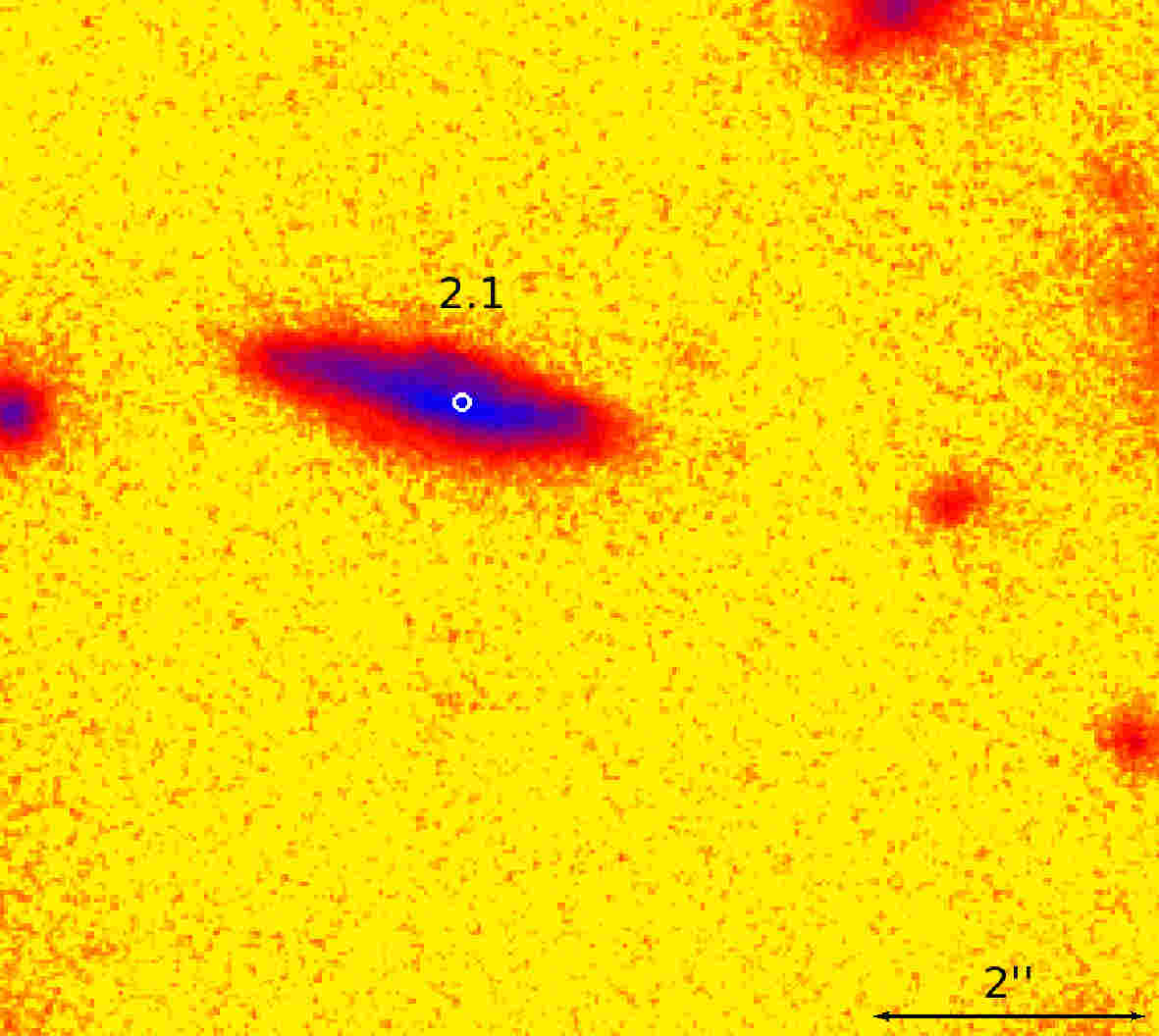}
        \caption{}
    \end{subfigure}
    \begin{subfigure}{0.49\textwidth}
        \centering
        \includegraphics[width=\textwidth]{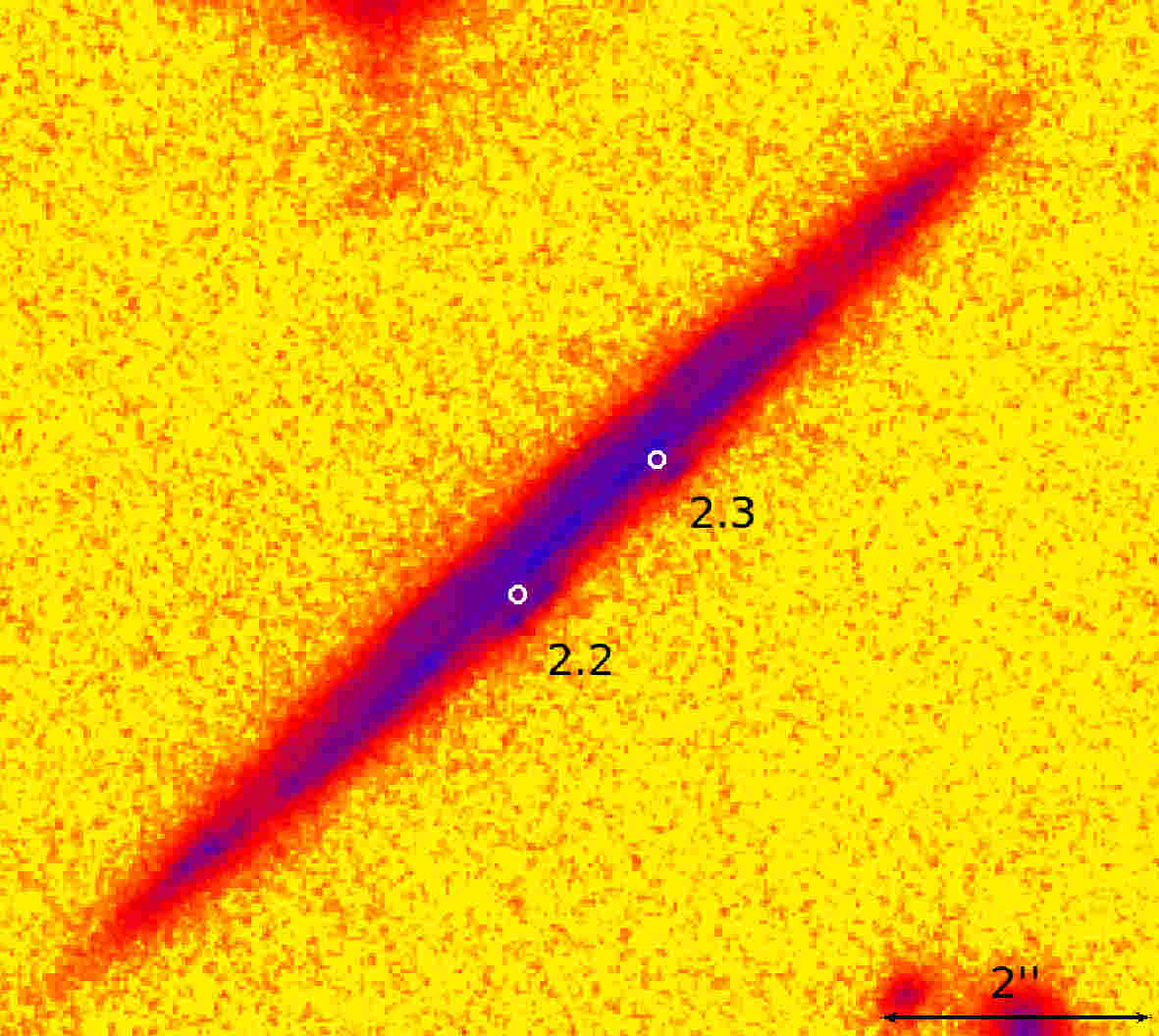}
        \caption{}
    \end{subfigure}
    \caption{Zoom-in on multiply lensed images of multiple image system  
             2. \textbf{(a)} Image 2.1. \textbf{(b)} Images 2.2 and 2.3. 
             \\ Due to the faintness of the arcs we show a false colour 
             version of the CLASH ACS-IR detection image (cf. Sect. 
             \ref{sec:data}).}
    \label{fig:sys2_arcs}
\end{minipage}
\end{figure*}

In Fig.~\ref{fig:sys2_arcs} we show multiple image system 2 which was already identified in \citet{Bradac08} and \citet{Halkola08} as well. It consists of one bright arc south-west of the cluster centre and one fainter arc south-east of the centre. However, we interpret the brighter arc to consist of two directly merging images on a tangential critical line due to the general straight elongated shape and especially due to substructures that seem to be exactly mirrored (cf. Fig.~\ref{fig:sys2_arcs}). This new interpretation has significant consequences for the entire mass distribution of the cluster and will be discussed further in Sect. \ref{sec:model_gl}. \citet{Halkola08} determined the spectroscopic redshift of this system to be $z_{spec}=1.75$. This is also consistent with the photometric redshift estimate from the CLASH ACS-IR catalogue ($z_{phot} = 1.78^{+0.01}_{-0.01}$).

\subsubsection{Multiple image system 3 (8 images + 1)}
\label{sec:img_sys3_data}

\begin{figure*}    
    \centering
    \begin{minipage}{140mm}
        \begin{subfigure}{0.33\textwidth}
            \centering
            \includegraphics[width=\textwidth]{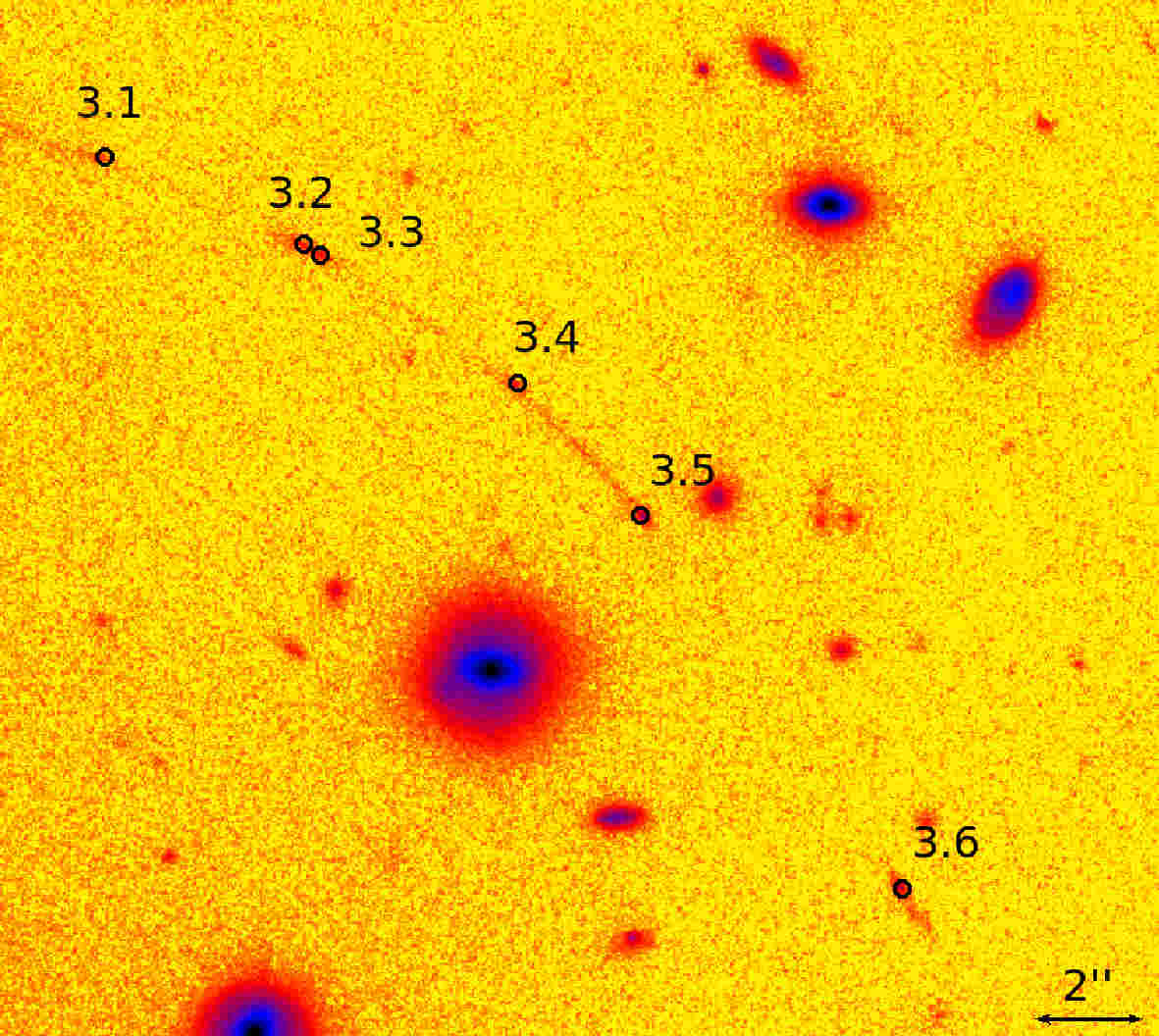}
            \caption{}
        \end{subfigure}
        \begin{subfigure}{0.33\textwidth}
            \centering
            \includegraphics[width=\textwidth]{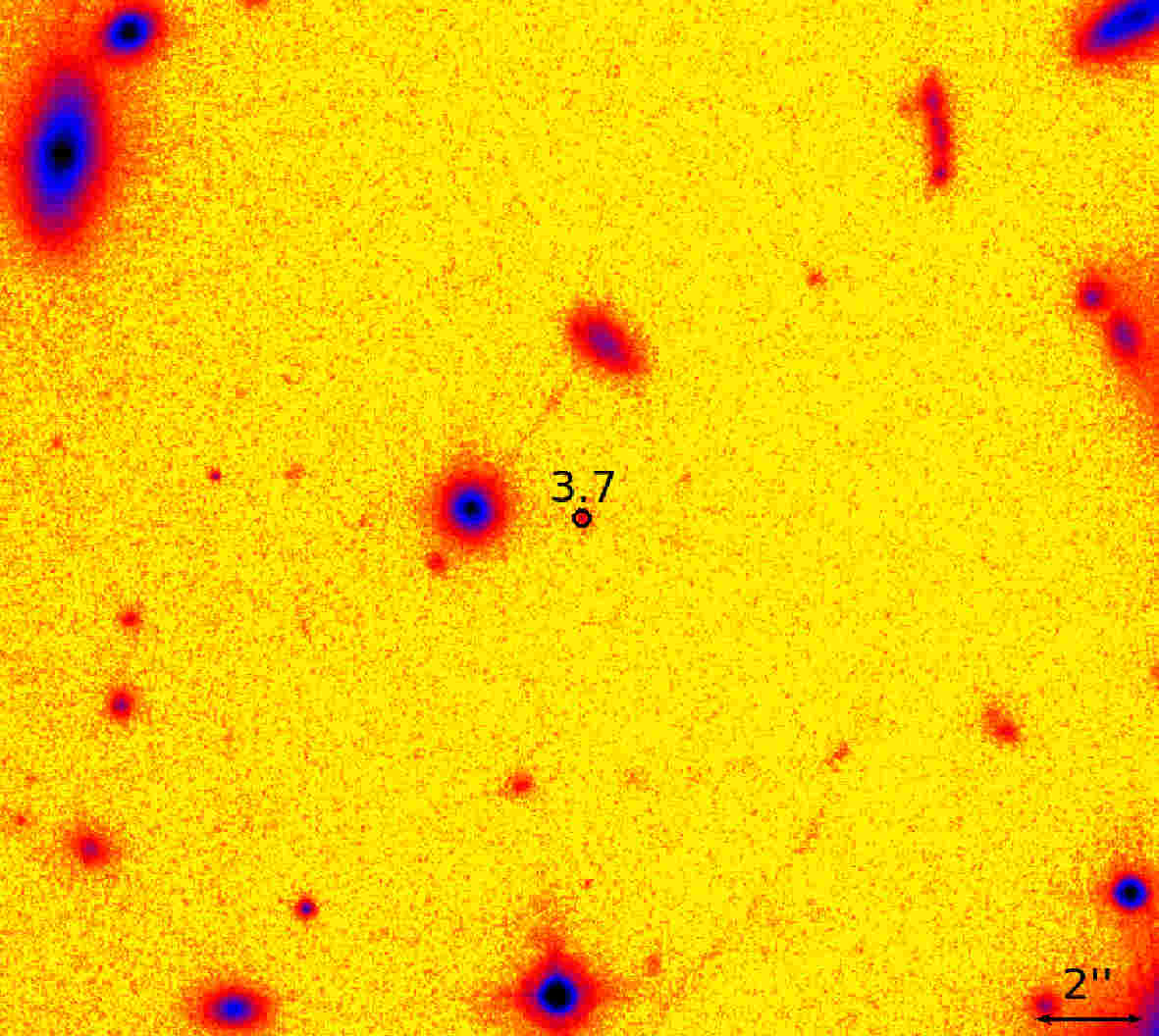}
            \caption{}
        \end{subfigure}
        \begin{subfigure}{0.33\textwidth}
           \centering
           \includegraphics[width=\textwidth]{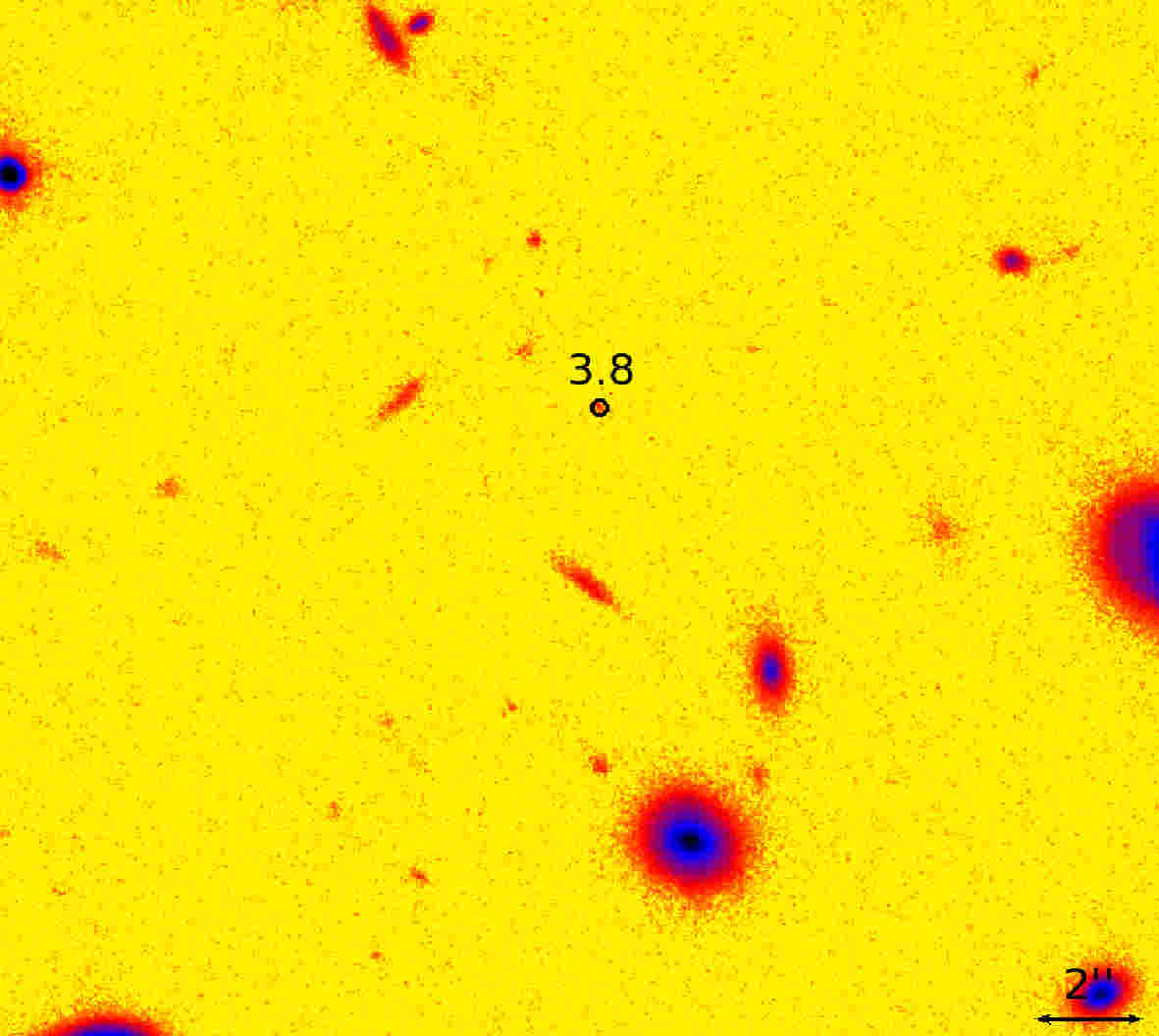}
           \caption{}
        \end{subfigure}
        \caption{Zoom-in on multiply lensed images of multiple image 
                 system 3. \textbf{(a)} Images 3.1 to 3.6. \textbf{(b)} 
                 Image 3.7. \textbf{(c)} Image 3.8. \\ Due to the 
                 faintness of the arcs we show a false colour version of 
                 the CLASH ACS-IR detection image (cf. Sect. 
                 \ref{sec:data}).}
        \label{fig:sys3_arcs}    
    \end{minipage}
\end{figure*}

The main part of this system (cf. Fig.~\ref{fig:sys3_arcs}) is located in the north-west of the cluster centre and is identified in \citet{Halkola08} as image systems 11 and 12, respectively. However, we interpret the system to originate from just one single source due to the consistent photometric redshifts of the images (cf. Tab.~\ref{tab:arcs}). Furthermore, we also include the two merging images (cf. images 3.2 and 3.3 in Fig.~\ref{fig:sys3_arcs}) that were not yet considered in previous studies. Nevertheless, their redshifts and location suggest their affiliation to the system, and we interpret them as two merging images on a tangential critical line. Altogether, this part of the system (cf. Fig.~\ref{fig:caust_img_sys3}) resembles strongly the massive arc structure in A370 (cf. \citealt{Richard09}).

Additionally to the north-western arc structure, there are two more images (cf. Fig.~\ref{fig:sys3_arcs}): one in the south-west and one in the south-east with respect to the cluster centre. Our model predicts further a ninth image directly in the cluster centre, which we were not able to detect yet due to confusion with the BCG.

The photometric redshift of the image system is derived from the CLASH ACS-IR catalogue to be $z_{phot}=4.19^{+0.12}_{-0.14}$. This redshift estimate is neither consistent with the photometric redshift estimates presented in \citet{Halkola08} ($z^{11a}_{phot} = 2.94 \pm 0.23$, $z^{11b}_{phot} = 3.61 \pm 0.20$, $z^{11c?}_{phot} = 2.80 \pm 0.73$ and $z^{12a}_{phot} = 2.79 \pm 0.75$, $z^{12c?}_{phot} = 1.75 \pm 1.09$; cf. Tab.~\ref{tab:arcs} for nomenclature) nor with the estimates from fitting strong-lensed data ($z^{B}_{fit} = 1.2 \pm 0.1$; cf. Tab.~\ref{tab:arcs} for nomenclature) by \citet{Bradac08}. 

Further implications of this system will be discussed in Sect. \ref{sec:img_sys3}.  

\subsubsection{Multiple image system 4 (3 images)}

\begin{figure*}
   \centering   
   \begin{minipage}{140mm}
       \begin{subfigure}{0.49\textwidth}
           \centering
           \includegraphics[width=\textwidth]{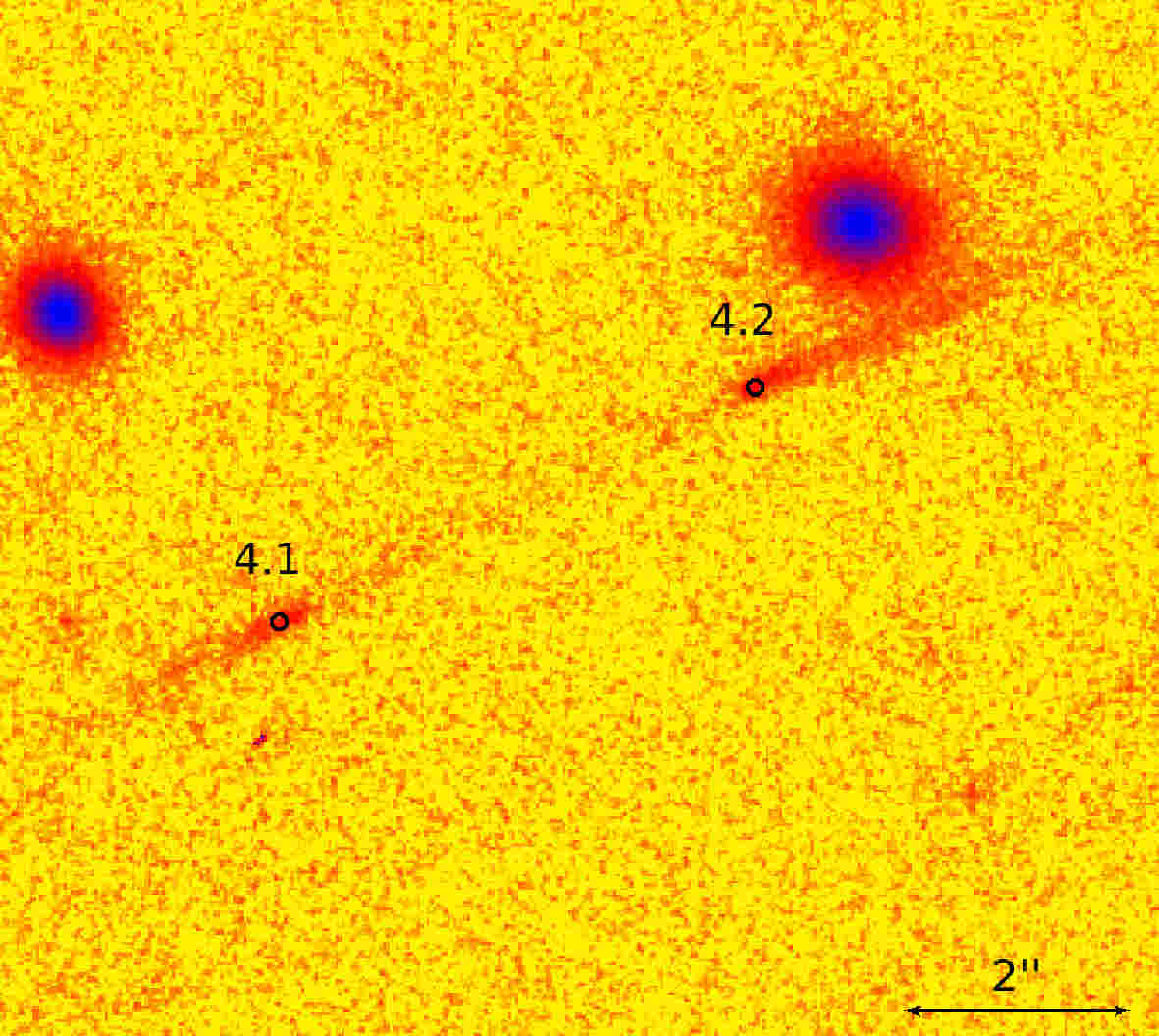}
           \caption{}
       \end{subfigure}
       \begin{subfigure}{0.49\textwidth}
           \centering
           \includegraphics[width=\textwidth]{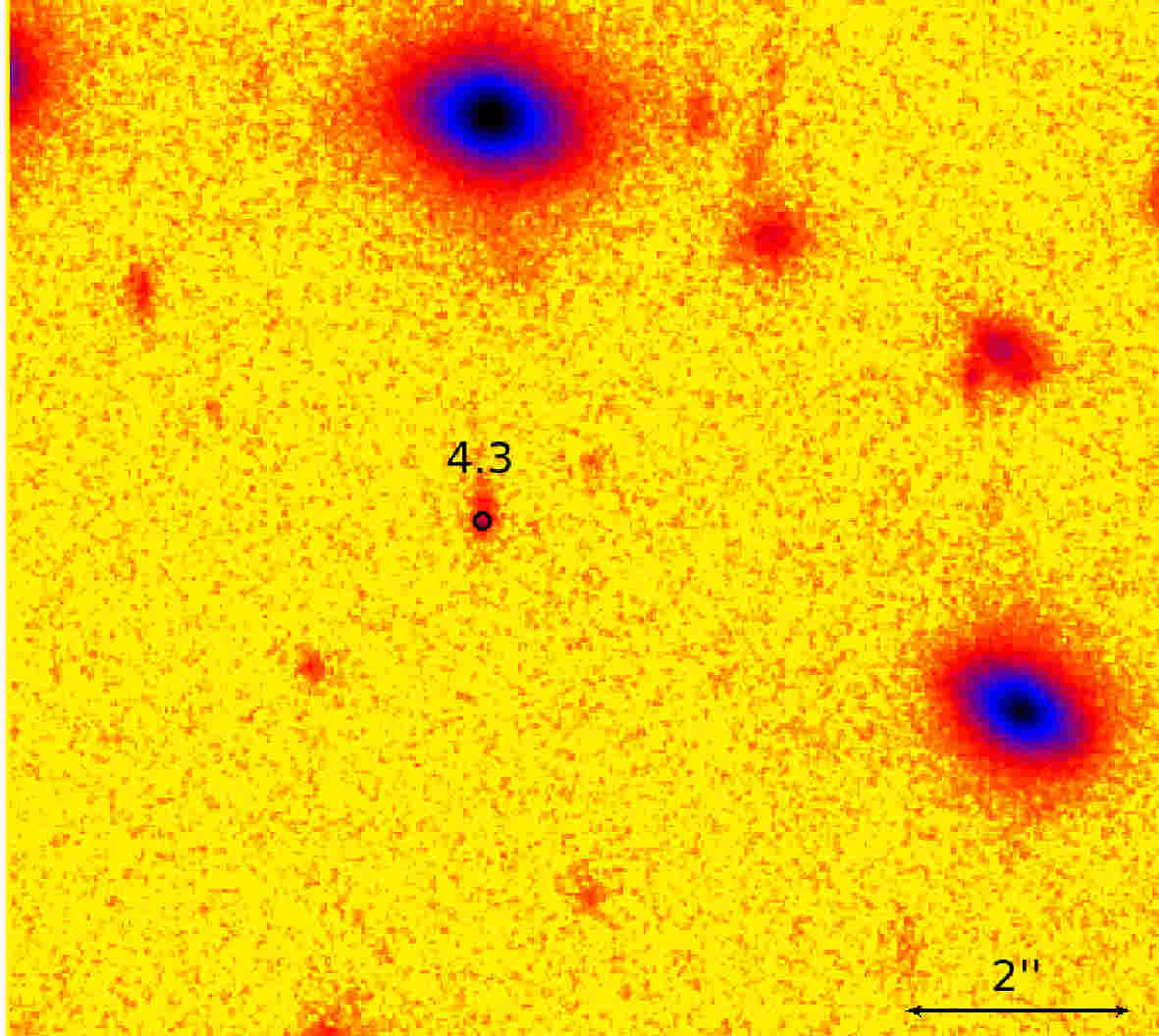}
           \caption{}
       \end{subfigure}
       \caption{Zoom-in on multiply lensed images of multiple image 
                system 4. \textbf{(a)} Image 4.1. \textbf{(b)} Images 
                4.2 and 4.3. \\ Due to the faintness of the arcs we show 
                a false colour version of the CLASH ACS-IR detection 
                image (cf. Sect. \ref{sec:data}).}
       \label{fig:sys4_arcs}
   \end{minipage}
\end{figure*}

This system (cf. Fig.~\ref{fig:sys4_arcs}) is identified in \citet{Halkola08} as system 8 consisting of a two-image system merging on a tangential critical line, and we confirm the third image (image 4.3) of this system predicted by \citet{Halkola08} as ''8c$?$``. In \citet{Bradac08} the two merging images are identified as well. All images are located in the north-east with respect to the centre of the cluster. Our model predicts the position of the third image (image 4.3) as well. We checked the redshifts of all images in the CLASH ACS-IR catalogue, and found them to be consistently around $z_{phot}=3.63^{+0.11}_{-0.08}$, which is also consistent with the photometric redshift estimate of image ''8c$?$`` ($z^{8c?}_{phot} = 3.68 \pm 0.11$; cf. Tab.~\ref{tab:arcs} for nomenclature), but not consistent with the redshift of image ''8a`` ($z^{8a}_{phot} = 1.88 \pm 0.19$; cf. Tab.~\ref{tab:arcs} for nomenclature) both estimated by \citet{Halkola08}. Neither is it consistent with the redshift estimate from fitting strong-lensed data ($z^{C}_{fit} = 2.0 \pm 1.0$; cf. Tab.~\ref{tab:arcs} for nomenclature) by \citet{Bradac08}. 

\section{Methods}
\label{sec:methods}

The analysis of strong lensing data can be performed either with parametric or non-parametric modelling software. For the parametric approach a certain (physically or observationally motivated) mass distribution has to be assumed already a priori. The analysis then consists of finding appropriate values for the parameters of the initially assumed mass model. In contrast to that, non-parametric approaches do not require any initial assumptions about the mass distribution. Note, however, the regularization is done as described below.

\subsection{Parametric -- \gl}
\label{sec:method_gl}

For the parametric mass reconstruction we have used the publicly available software \gl\footnote{We have used version 1.1.5 for our analysis. The software can be downloaded from: \url{http://www.slac.stanford.edu/~oguri/glafic/}} \citep{Oguri10}. Similar to other parametric software packages like \texttt{LENSTOOL} \citep{Kneib96, Jullo07, JulloKneib09}, \gl\ offers to set up a multiple component mass model where each component can be described by a large variety of different density profiles. These profiles are usually defined by six to seven parameters (e.g. velocity dispersion, orientation angle and so on).\\

The final parametric model we used to obtain the mass estimates presented further below consists of two smooth mass components for the dark matter including the mass of the ICM, two profiles for the cD galaxies, profiles for further cluster members and finally two additional perturbers (cf. Sect. \ref{sec:img_sys3}). In the following we will discuss the particular types of profiles used for each component in more detail.

\subsubsection{Cluster galaxies \& perturbers}
\label{sec:jaffe}

For modelling the mass distribution of cluster member galaxies we employ a pseudo-Jaffe ellipsoid (\texttt{jaffe} in \gl) whose three-dimensional radial density profile is given by 
\begin{equation}
  \rho (r) \propto \frac{\sigma}{(r^2+r_{core}^2)(r^2+r_{trunc}^2)} \, , 
\end{equation}
with velocity dispersion $\sigma$, core radius $r_{core}$ and truncation radius $r_{trunc}$.
We apply this profile in particular for modelling both cD galaxies individually, as well as both perturbers. In contrast to that the parameters of the other remaining cluster member galaxies are linked according to the following scaling relations
\begin{equation} \label{eq:scaling_rel1}
 \frac{\sigma_i}{\sigma_*} = \left( \frac{L_i}{L_*} \right)^{\frac{1}{4}} \ \mathrm{and} \ \ 
 \frac{r_{trunc,i}}{r_{trunc,*}} = \left( \frac{L_i}{L_*} \right)^{\frac{1}{2}} \, ,
\end{equation}       
thus yielding a constant mass-to-light ratio $M/L$ \citep{NatarajanKneib97}. This approach was also performed by \citet{Bradac08} and \citet{Halkola08}.

The luminosities $L_i$ were derived from the F$814$W magnitudes taken from the provided CLASH ACS-IR catalogue as well as the other parameters needed. Only the position angles $\theta_i$ are not provided in the catalogue, but can be easily estimated from a \texttt{SExtractor} \citep{Bertin96} run on one of the respective detection images. The reference values for $L_*$, $\sigma_*$ and $r_{trunc, \, *}$ were chosen such that a galaxy with $mag^{F814W}_{AB} = 20.5$ has a velocity dispersion $\sigma = 260 \, \mathrm{km \, s}^{-1}$ and a truncation radius $r_{trunc} = 5 \,$kpc which is the same normalization as given in \citet{Bradac08}.

This catalogue of cluster member galaxies contains $24$ out of the $48$ spectroscopically confirmed cluster members from \citet{Lu10} that are within radius $R \leq 75''$ from the cluster centre. Additionally, we also included the brightest galaxies that are also within radius $R \leq 75''$ from the cluster centre with redshifts in the range $0.4 \leq z_{gal} \leq 0.5$ and that are not already contained in the previous sample of $24$ galaxies. In total, we compiled a catalogue (cf. Tab.~\ref{tab:cluster_members}) of $101$ cluster member galaxies.

\begin{table*}
\caption{Cluster member galaxies as used for our analysis in \gl . Positions are given with respect to the cluster centre at position $\text{RA}=206.8775 \, \deg$, $\text{Dec}=-11.7526 \, \deg$ (J$2000$). Spectroscopically confirmed cluster members by \citet{Lu10} are marked with ''$\ast$ ``. The complete table is available online.}
\label{tab:cluster_members}
\centering
\begin{tabular}{rrrrrc}
\hline
\multicolumn{1}{c}{$\Delta$RA$ \ ['']$} & \multicolumn{1}{c}{$\Delta$Dec$ \ ['']$} & \multicolumn{1}{c}{$L^{F814W}_i / L^{F814W}_*$} & \multicolumn{1}{c}{$e$} & \multicolumn{1}{c}{$\theta \ [\deg]$} & \citet{Lu10} \\ 
\hline
-39.7217 & 48.1187 & 3.53737 & 0.288 & 76.6772 &  \\ 
17.8258 & -53.1457 & 2.94307 & 0.251 & 20.6834 &  \\ 
41.5544 & -44.2282 & 2.53396 & 0.206 & -42.7160 & $\ast$ \\ 
50.8514 & -33.4127 & 2.46740 & 0.275 & -87.0736 & $\ast$ \\ 
-27.2336 & 79.3580 & 2.24430 & 0.362 & 24.1564 & $\ast$ \\
\multicolumn{1}{c}{...} & \multicolumn{1}{c}{...} & \multicolumn{1}{c}{...} & \multicolumn{1}{c}{...} & \multicolumn{1}{c}{...} & \multicolumn{1}{c}{...} \\
\end{tabular}
\end{table*}

\subsubsection{Smooth mass component}
\label{sec:model_gl}

Considering the merger history of the cluster as mentioned in Sect. \ref{sec:intro} and the studies from \citet{Halkola08} and \citet{Bradac08}, we introduced two smooth cluster components in our parametric model right from the beginning as well. Although the topic is still under discussion, one of the most appropriate types of profiles for the dark matter dominated smooth cluster component seems to be the Navarro--Frenk--White (NFW) profile \citep{Navarro97}. This form of universal mass halo is predicted by cosmological dark matter simulations. Its three-dimensional radial density profile is given as
\begin{equation}
  \rho_{nfw}(r) = \frac{\rho_s}{(r/r_s)(1+r/r_s)^2} \, ,
\end{equation}
with the characteristic density $\rho_s$ and the scale radius $r_s$.

The \textit{concentration parameter} $c$ is defined as the ratio of the virial radius $r_{vir}$ to the scale radius $r_s$,
\begin{equation} \label{eq:c}
c = \frac{r_{vir}}{r_s} \, .
\end{equation}
\citet{Oguri10} defines the virial mass $M$ in \gl\ as
\begin{equation} \label{eq:M}
M = \frac{4 \pi}{3} r^3_{vir} \Delta(z) \overline{\rho}(z) = \int_0^{r_{vir}} 4 \pi r^2 \rho_{nfw}(r) \, \mathrm{d}r \, ,
\end{equation}
where the expression $\Delta(z) \overline{\rho}(z)$ describes the mean overdensity inside a sphere with radius $r_{vir}$. The nonlinear overdensity $\Delta(z)$ is evaluated by adopting the fitting formula of \citet{NakamuraSuto97}.

\subsubsection{Optimization and uncertainties}
\label{sec:errors_gl}

The optimization of the parameters of the assumed mass profiles is based on a $\chi^2$-minimization with a \textit{downhill-simplex} algorithm \citep{NelderMead65}, described in more detail in \citet{Oguri10}.
Due to the complexity of the mass model we restrict all calculations to the source plane, where we approximate the $\chi^2$ of the $i$th image per multiple image system as \citep{Oguri10}
\begin{equation}
 \chi^2_{pos} \approx \chi^2_{pos, \, src} = \sum_i \frac{(\boldsymbol{u}_{i, \, obs}-\boldsymbol{u})^T \boldsymbol{M}_i^2 (\boldsymbol{u}_{i, \, obs}-\boldsymbol{u})}{\sigma_{i, \, pos}^2} \, .
\end{equation} 
This 'corrected' source-plane $\chi^2$ follows from the assumption that the fitted image plane position is close to the observed image position.
Then, the magnification tensor $\boldsymbol{M}_i = \mathrm{d} \boldsymbol{x}_i/ \mathrm{d} \boldsymbol{u}$ can be used to approximately relate the image positions $\boldsymbol{x}$ to the source-plane positions $\boldsymbol{u}$ (\citealt{Kochanek91}), i.e.
\begin{equation}
\label{eq:approximation sp}
 \boldsymbol{x}_{i, \, obs} - \boldsymbol{x}_i \approx \boldsymbol{M}_i (\boldsymbol{u}_{i, \, obs}-\boldsymbol{u}) \, .  
\end{equation}

We set the positional uncertainties $\sigma_{i, \, pos}$ of multiply lensed images to $0.5''$ which is higher than the typically observed value of $\sim 0.1''$ for measurements by HST (e.g. \citealt{Golse02}). However, setting the uncertainties that low during the optimization process was too restrictive for the optimization routine. This is in fact necessary because of the unknown matter distribution along the line-of-sight, which may change positions by this amount on the scale of galaxy clusters (M. Bartelmann, priv. comm.).

For the estimation of uncertainties of model parameters, \gl\ provides built-in functions in order to perform a Bayesian likelihood interpretation employing a Monte Carlo Markov Chain (MCMC) approach. However, a non-trivial complication arises, due to the limitation of all calculations to the source plane. Although \gl\ also provides routines in order to perform MCMC calculations in the source plane in general, a test run of the built-in MCMC routine of \gl\ in the source plane revealed that a fair fraction of the models did not correspond to the observations, for example, predicting an incorrect amount of images per multiple image system or incorrect image positions. The reason is that equation~\ref{eq:approximation sp} is not valid in the direct vicinity of a caustic. In regions of strongly clustered caustics this leads sometimes to wrong models, which we needed to discard.

Since a direct inversion for each single $\chi^2$--calculation is too time consuming (e.g. one sequence of $\chi^2$-minimizations of the mass model presented in Sect. \ref{sec:models_gl} in the image plane takes about one month on a typical work station\footnote{Four CPUs with $2.66 \,$GHz each and $8 \,$GB RAM. Note that \gl\ is not parallelized yet.}), we employed a Monte Carlo approach, i.e. we varied the input data within the given uncertainties in order to derive uncertainties for the model parameters including a check for the correct amount and positions of images at the end of a full $\chi^2$-minimization run. In particular, the input data for \gl\ consists of the positions of the images in the multiple image systems (cf. Tab.~\ref{tab:arcs}). So, the initial positions were varied by drawing random positions from a Gaussian distribution centred on the observed position with width $\sigma = 0.5''$ corresponding to the assumed positional uncertainties of the images. Furthermore, the positions of merging images were linked to each other, so that they were always shifted by the same amount in the same direction in order to avoid shifts of the images against each other (positional fluctuations due to large scale structure affect larger scales). 

For each set of varied input data we performed a $\chi^2$-minimization run of the mass model in the source plane based on the initial best-fitting parameters. The photometric redshifts for multiple image systems 1, 3 and 4 were also free to vary within the given errors. In total $500$ models were calculated. Finally, the prediction of the correct amount of images per multiple image system and their positions were checked visually for each such model by performing a full inversion of the lens equation. This resulted in $366$ accepted models (i.e. $\approx 73 \, \%$) from which the $68 \, \%$ confidence intervals for the results obtained with \gl\ were calculated.

Since the optimization for a set of varied input data started always from the best-fitting model parameters, this sampling method is rather insensitive to possibly existing, entirely different solutions for the mass model, and assumes implicitly that the best-fitting model is the global minimum in solution space. 
 
\subsection{Non-parametric -- \pl}
\label{sec:method_pl}

Detailed information about the functionality of \pl\footnote{We have used version 2.17 for our analysis. The software can be downloaded from: \url{http://www.qgd.uzh.ch/projects/pixelens/}} is presented in \citet{Saha97, Saha04}. The non-parametric approach in \pl\ employs a formulation of the lens equation in terms of the arrival-time surface. Introducing square mass pixels allows it then to express the effective lensing potential in terms of the convergence $\kappa$ such that the lens equation becomes linear in the unknowns, $\kappa$ and source plane positions $\boldsymbol{\beta}$.  
Observations of image positions of multiply lensed systems, and in general also time delay information, then impose constraints on the linear equation system.

In addition to restricting all calculations again to the source plane, such an equation system remains under-determined which results in a whole family of best-fitting models for a given image configuration.

In order to deal with this situation, \pl\ employs a built-in MCMC approach and creates an ensemble of $100$ lens models per given image configuration. Since all equations are linear in the unknowns, the best-fitting model and its uncertainties are obtained by finally averaging over the ensemble.

Note that in addition to the standard regularization assumptions of \pl\ (''the prior``, cf. \citealt{Saha04}), we demand further that the tilt of isocontours is $\leq 60 \deg$. 

\section{Analysis}
\label{sec:analysis}

\subsection{Parametric -- \gl} 
\label{sec:models_gl}

For the parametric modelling, we did not include all image systems available and all their respective images at once. Instead, we included them iteratively. Thus, deriving the best-fitting model in \gl\ was a sequence of including a multiple image system with subsequent source plane optimization of the proposed mass distribution, and then checking the predictions via a full inversion of the lens equation in order to calculate predicted images. If additional images are predicted by the mass model, these have to be checked carefully by comparing their position, morphology, colour, surface brightness and redshift with all other confirmed images of the multiple image system. Here especially, the CLASH catalogues constitute an invaluable data source and facilitate the decision whether or not to include or refuse such an additionally predicted image, and hence improve the whole mass model significantly. 

Based on the F$814$W filter, we define a limiting magnitude for a $5\sigma$ source detection as $m^{F814W}_{lim} = ZP^{F814W} - 2.5 \log_{10}(5 \sqrt{N_{pix}}\sigma_{bkg})$ (e.g. \citealt{Erben09}), where $ZP$ is the extinction-corrected magnitude zeropoint in F$814$W, $N_{pix}$ the minimal number of continuous pixels that define a source in the catalogue (i.e. $N_{pix}=9$) and $\sigma_{bkg}$ the sky background noise estimation. 
The values for the limiting magnitude and surface brightness are then $m^{F814W}_{lim} \approx 28.4 \, \mathrm{mag}$ and $S^{F814W}_{lim} \approx 24.9 \, \mathrm{mag/arcsec^2}$, respectively.
   
If the additional image is indeed a correct prediction (with a candidate source) with respect to the limits defined above, it will be included as a new constraint for the next optimization step in addition to the next multiple image system. If the additional image is a false prediction (no candidate source available), one will vary the input model parameters and start with those a new optimization run.

Eventually, this whole process converged to a best-fitting model including all four multiple image systems presented in Sect. \ref{sec:mult_img}. These provide in total $38$ constraints from observed image positions with which the $28$ free model parameters were fit, yielding a reduced $\chi^2_{red}$ of $0.68$ ($9$ degrees of freedom). We did not include further constraints from fluxes as a measure for the magnification ratio, since the systematic errors for flux constraints are still under discussion (e.g. \citealt{Kochanek91}, \citealt{Liesenborgs12}).\\

\begin{figure*}
	\centering
	\begin{minipage}{180mm}
    \begin{subfigure}{0.49\textwidth}
        \centering        
        \includegraphics[width=\textwidth]{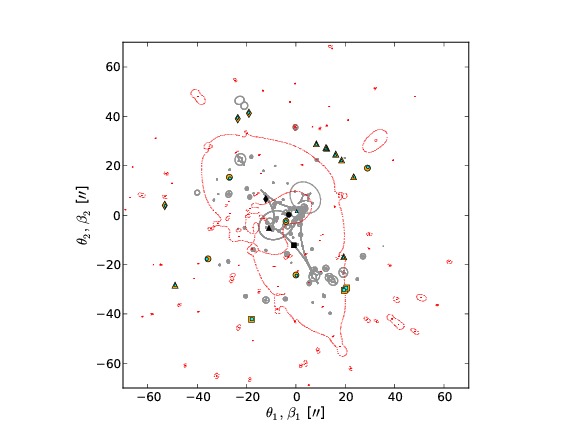}
        \caption{}
    \end{subfigure}
    \begin{subfigure}{0.49\textwidth}
        \centering
        \includegraphics[width=\textwidth]{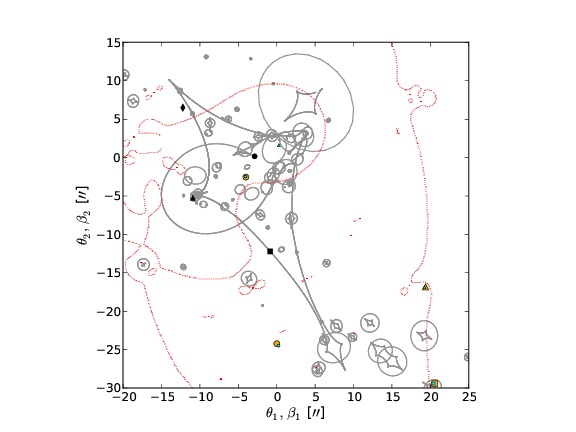}
        \caption{} 
    \end{subfigure}
    \caption{Critical curves (red, dotted lines) and caustics (grey, 
             solid lines) in RX J1347.5-1145 obtained from the best- 
             fitting model in \gl . The critical curves and caustics are 
             plotted for the redshift of multiple image system 2 at $z = 
             1.75$. Furthermore, we show observed images (orange), 
             predicted images (cyan) and the respective sources (black) 
             for all four multiple image systems. Different symbols 
             denote different multiple image systems (system 1: 
             $\bigcirc$; system 2: $\Box$; system 3: $\triangle$; system 
             4: $\diamondsuit$). \textbf{(a)} Total area used for the 
             strong lensing analysis. \textbf{(b)} Zoom-in on central 
             part.}
        \label{fig:best_fit}
    \end{minipage}	 
\end{figure*}

The critical curves and caustics with observed and predicted image positions for all multiple image systems included in the best-fitting model are presented in Fig.~\ref{fig:best_fit}. 

In Tables \ref{tab:best_nfw}, \ref{tab:best_jaffe} and \ref{tab:perturbers} we present an overview of the fitted values of parameters of the best-fitting model in \gl\ including the $68 \, \%$ confidence intervals estimated from the Monte Carlo sampling as described in Sect. \ref{sec:errors_gl}. 

\begin{table*}
\caption{Parameters for the smooth mass components described by NFW profiles from the best-fitting model in \gl , the uncertainties are estimated as described in Sect. \ref{sec:errors_gl}. ''NFW1`` is located close to the BCG and ''NFW2`` is in the south-east of the cluster (cf. Fig.~\ref{fig:kappa_gl}). For a detailed explanation of the parameters and model please refer to Sect. \ref{sec:model_gl}.}
\label{tab:best_nfw}
\begin{center}
\begin{tabular}{ccccccc}
\hline
\parbox[0pt][1.5em][c]{0cm}{} Model & $M \ [\mathrm{\Msol} \, h^{-1}]$ & $\Delta$RA$ \ ['']$ & $\Delta$Dec$ \ ['']$ & $e$ & $\theta \ [\deg]$ & c \\ \hline
\parbox[0pt][1.5em][c]{0cm}{} NFW1 & $(5.75^{+0.72}_{-0.35}) \times 10^{14}$ & $0.14^{+0.36}_{-0.45}$ & $5.15^{+0.59}_{-0.34}$ & $0.15^{+0.04}_{-0.02}$ & $-144.91^{+2.11}_{-4.63}$ & $6.08^{+0.18}_{-1.15}$ \\ 
\parbox[0pt][1.5em][c]{0cm}{} NFW2 & $(5.23^{+0.39}_{-0.56}) \times 10^{14}$ & $-8.26^{+0.41}_{-0.50}$ & $-12.84^{+0.43}_{-0.68}$ & $0.66^{+0.02}_{-0.03}$ & $-145.80^{+0.70}_{-0.74}$ & $4.72^{+0.24}_{-0.24}$ \\ \hline
\end{tabular}
\end{center}
\end{table*}

\begin{table*}
\caption{Parameters for the two cD galaxies in the cluster each modelled separately with a pseudo-Jaffe profile (cf. Sect. \ref{sec:jaffe} also for a detailed explanation of the parameters) from the best-fitting model obtained in \gl .}
\begin{center}
\begin{tabular}{cccccccc}
\hline
\parbox[0pt][1.5em][c]{0cm}{} Object & $\sigma \ [\mathrm{km} \, \mathrm{s}^{-1}]$ & $\Delta$RA$ \ ['']$ & $\Delta$Dec$ \ ['']$ & $e$ & $\theta \ [\deg]$ & $r_{trunc} \ ['']$ & $r_{core} \ ['']$ \\ \hline
\parbox[0pt][1.5em][c]{0cm}{} BCG & 593.59 & 0.00 & 0.00 & 0.32 & 2.16 & 4.89 & 0.95 \\
\parbox[0pt][1.5em][c]{0cm}{} $2^\mathrm{nd}$ cD & 344.00 & -18.27 & -1.95 & 0.26 & 44.37 & 3.87 & 0.79 \\ \hline
\end{tabular}
\end{center}
\label{tab:best_jaffe}
\end{table*}

\begin{figure*}
    \centering
    \begin{minipage}{140mm}
         \begin{subfigure}{0.49\textwidth}
             \centering
             \includegraphics[width=\textwidth]{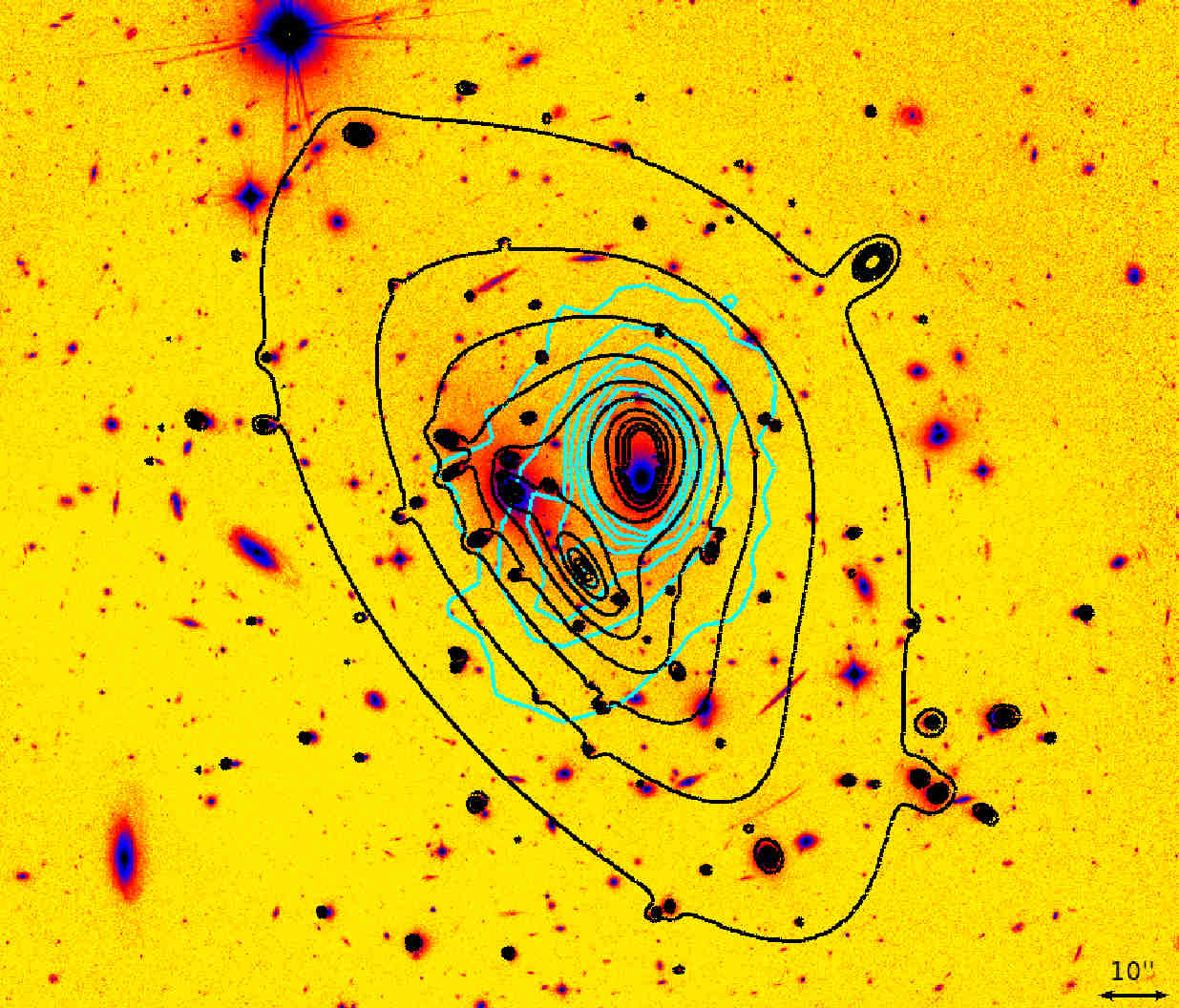}
             \caption{}
         \end{subfigure}
         \begin{subfigure}{0.49\textwidth}
              \centering              
              \includegraphics[width=\textwidth]{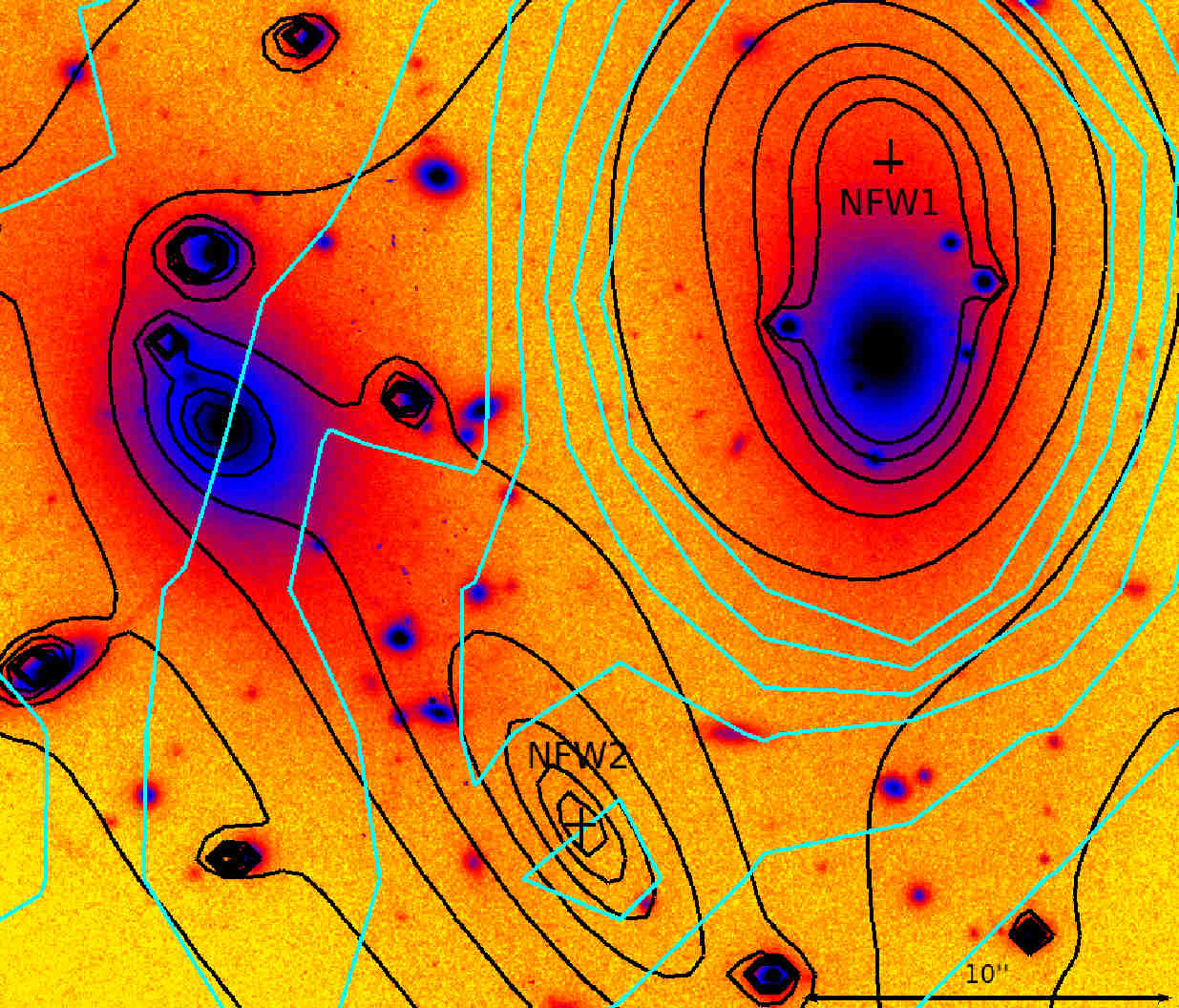}
              \caption{}
         \end{subfigure}
         \caption{\textbf{(a)} Surface mass density contours (solid, 
                  black lines) in units of the critical surface mass 
                  density obtained from the best-fitting parametric 
                  model in \gl\ for redshift $z=1.75$. Furthermore, we 
                  show X-ray brightness contours (solid, cyan lines) 
                  from \textit{Chandra}. The contours of the second NFW 
                  profile (NFW2) coincide with the centre of the south-
                  eastern extension of the X-ray surface              
                  brightness contours. \textbf{(b)} Zoom in on the 
                  positions of the NFW profiles (black crosses, showing 
                  the $68 \, \%$ confidence interval for the position), 
                  mass and X-ray contours.}
         \label{fig:kappa_gl}
    \end{minipage}
\end{figure*}

Finally, we show in Fig.~\ref{fig:kappa_gl} the convergence contours obtained from the best-fitting model. The contours are very elliptical and resemble the contours obtained with \pl\ (cf. Fig.~\ref{fig:mass_prim_sec_pl}) in general. 

The distinct ellipticity may be caused by the new interpretation of the prominent, elongated arc feature (images 2.2 and 2.3) of multiple image system 2 (cf. Fig.~\ref{fig:sys2_arcs} and Sect. \ref{sec:mult_img}) to count as two merging images, and not just as one image constraint, as it was described in both \citet{Bradac08} and \citet{Halkola08}\footnote{\citet{Halkola08} present a simulated image of images 2.2 and 2.3 based on their best-fitting model which does suggest they also modelled it as two merging images.}. However, when counting this elongated arc as the result of two merging images, the course of the critical curve in this region must necessarily change such that it must go straight through the symmetry axis of this arc (cf. Fig.~\ref{fig:best_fit}). This in turn requires an adjustment of the mass profile accordingly, and this was achieved using \gl\ by shifting the second NFW profile away from the position of the second cD galaxy (where we had placed the second NFW halo initially) towards the south-west. It was necessary to keep the position of the second NFW profile free [not fixed at the position of the second cD galaxy], as we already mentioned in Sect. \ref{sec:jaffe}. 

\subsubsection{Multiple image system 3 in detail} 
\label{sec:img_sys3}

Apart from the new interpretation of the elongated arc in multiple image system 2 and its consequences for the mass distribution, another new interpretation was to count images 3.1 to 3.8 as only one multiple image system. This interpretation was not at all clear from the beginning due to differing interpretations of the affiliation of these images in \citet{Bradac08} and \citet{Halkola08}. Initially, we followed the image affiliations presented in the latter study, since their definitions appeared to be more consistent. This means in particular that we also affiliated images 3.1 to 3.8 with two different multiple image systems (we want to emphasize again, that images 3.2 and 3.3 were not at all considered in both studies, although they were visible in earlier data sets) in the beginning of our analysis. 

However, whenever models were employed with system 3 split into two systems, \gl\ predicted additional images in the vicinity of images 3.7 (''11c$?$`` in \citealt{Halkola08}; also cf. Tab.~\ref{tab:arcs} for nomenclature) and 3.8 (''11d$?$`` in \citealt{Halkola08}), respectively. This is indeed expected from the small separation of images 3.1 to 3.6.

A search in the CLASH catalogues revealed no second image with appropriate redshift in the vicinity of image 3.8 (''11d$?$`` in \citealt{Halkola08}) and the proposed candidate image (''12c$?$``) from \citet{Halkola08} is included in the CLASH ACS-IR catalogue with a too low redshift of $z_{phot}=0.87^{+0.18}_{-0.36}$. 

\begin{figure*}
    \centering
    \begin{minipage}{140mm}
        \begin{subfigure}{0.49\textwidth}
            \centering      
            \includegraphics[width=\textwidth]{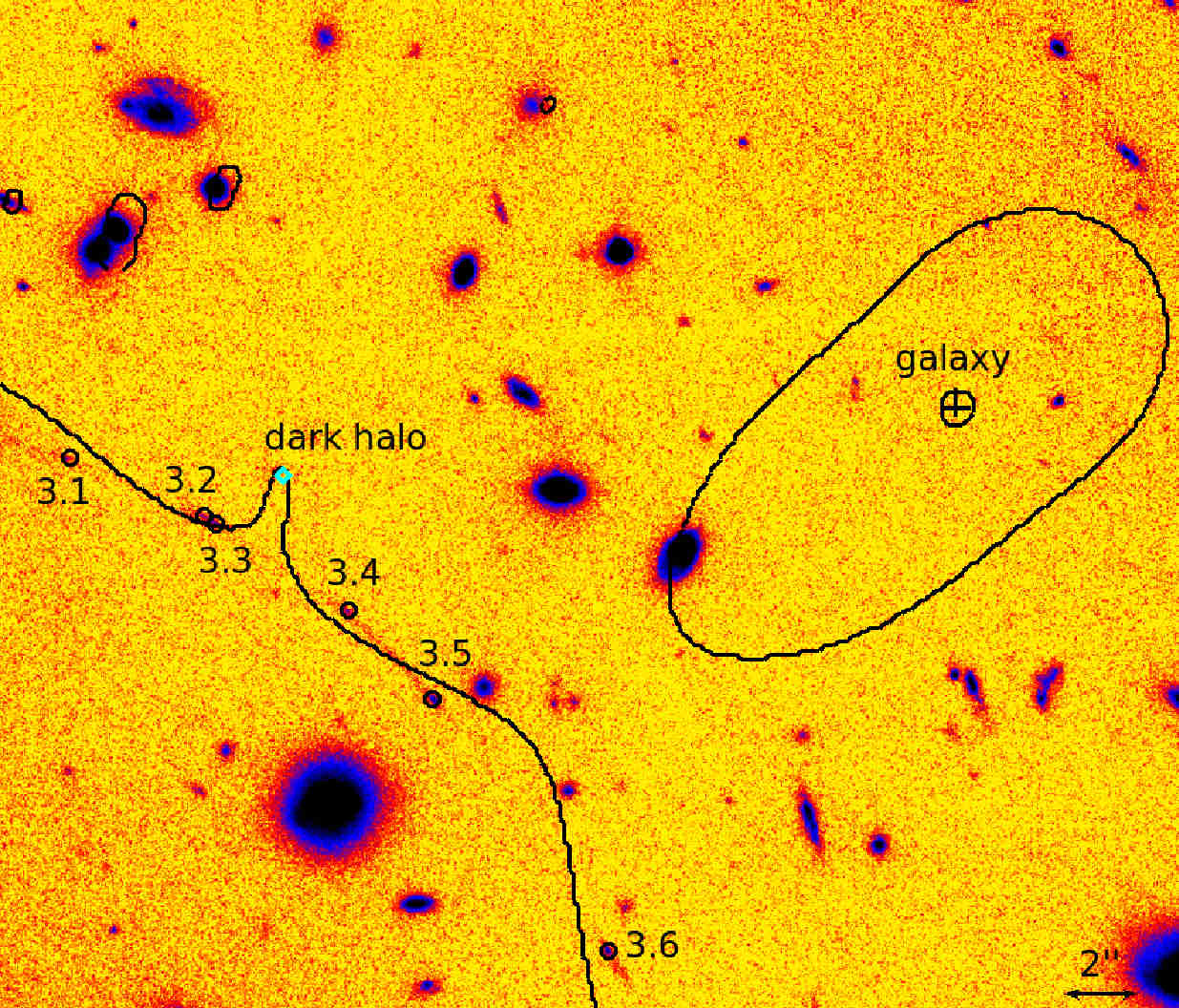}
            \caption{}
        \end{subfigure}
        \begin{subfigure}{0.49\textwidth}
            \centering
            \includegraphics[width=\textwidth]{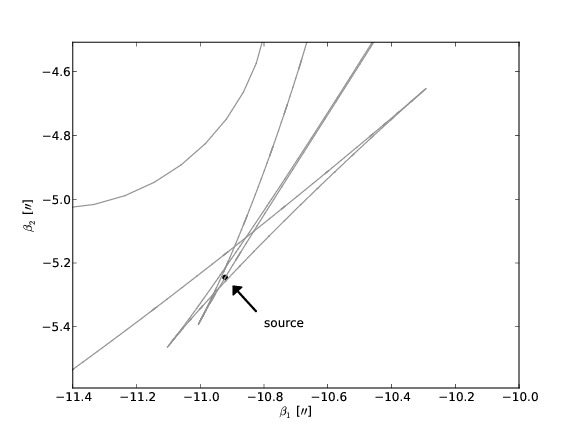}
            \caption{}
        \end{subfigure}
        \caption{\textbf{(a)} Critical curves (solid, black lines) close 
                 to images 3.1 to 3.6 at redshift $z = 4.19$ in RX 
                 J1347.5-1145. Furthermore, the positions of the 
                 perturbers necessary for this course of the critical 
                 curves are shown. Very close to the position of the 
                 object ''dark halo`` (cyan) is a very faint source 
                 visible, whereas the object ''galaxy`` (black cross) 
                 represents a massive model of a possible overdensity 
                 in this region. Due to the faintness of images 3.1 to 
                 3.6 we use a high contrast, false colour version of the 
                 CLASH ACS-IR detection image. \textbf{(b)} Multiply 
                 folded ''swallowtail`` caustics (solid, grey lines) and 
                 position of the ''source`` (black) for image system 3 
                 within these.\\ 
                 Compare to similar arc feature in A370 (\citealt{Richard09}).}
    \label{fig:caust_img_sys3} 
    \end{minipage}   
\end{figure*}
  
Therefore and further because of the redshifts of all images later found to be consistent due to the then available CLASH catalogues, we finally interpreted all images to originate from one source only. 

Moreover, the north-western arc structures (images 3.1 to 3.6 in Fig.~\ref{fig:caust_img_sys3}) of image system 3 strongly resemble the prominent arc feature of A370 (cf. \citealt{Richard09}), as mentioned in Sect. \ref{sec:mult_img}. The special arc configuration in A370 is caused by the source being located on a doubly folded caustic (a so-called ''swallowtail``). 
  
Guided by this resemblance between multiple image system 3 and the multiple image system in A370, we adopted this swallowtail folding by at first modelling one of the more luminous neighbouring galaxies separately at a fixed position, making use again of the pseudo-Jaffe profile (cf. Sect. \ref{sec:jaffe}). Later the coordinates were left free for fitting as well, because during the optimization high masses were assigned to this object (i.e. ''galaxy`` in Tab.~\ref{tab:perturbers}). Thus, we rather tend to interpret the object ''galaxy`` to represent an overdensity in the whole north-western part of the cluster than just to count it as an individual massive object. 

In addition to that, we further included another perturber (i.e. ''dark halo`` in Tab.~\ref{tab:perturbers}) in close vicinity to the two merging images 3.2 and 3.3 modelled again with a pseudo-Jaffe profile (cf. Sect. \ref{sec:jaffe} and Tab.~\ref{tab:perturbers}) in order to force the critical line to bend exactly through the merging images (cf. Sect. \ref{sec:img_sys3_data}). All seven parameters of this additional mass halo were left free for fitting (of course providing appropriate initial values, especially for its position). Interestingly, the fitted position for the additional profile coincides with the position of a very small and faint object (cf. Fig.~\ref{fig:caust_img_sys3}). A search for this object in both CLASH catalogues revealed a redshift limit of $z_{phot} > 0.7$ and a very high upper limit of $z \approx 3$. Although an unambiguous determination of its photometric redshift is not possible at the moment, the data rather suggest to assign the object to the background of the cluster ($z_{cluster} = 0.451$). Whether this really means that this additional mass halo is not physically related to the respective faint object at all and thus rather another dark matter overdensity, is hard to assess at the moment. See also \citet{Liesenborgs12} on the degeneracies involved in modelling cluster lens components.

\begin{table*}
\caption{Model parameters for the additionally predicted perturbers which were both modelled with the pseudo-Jaffe profile (cf. Sect. \ref{sec:jaffe} also for a detailed explanation of the parameters). In case the parameters were free for optimization during the error estimation process (cf. Sect. \ref{sec:errors_gl}) the respective $68 \, \%$ confidence intervals are noted as well.}
\label{tab:perturbers}
\begin{center}
\begin{tabular}{cccccccccc}
\hline
\parbox[0pt][1.5em][c]{0cm}{} Object & $\sigma \ [\mathrm{km} \, \mathrm{s}^{-1}]$ & $\Delta$RA$ \ ['']$ & $\Delta$Dec$ \ ['']$ & $e$ & $\theta \ [\deg]$ & $r_{trunc} \ ['']$ & $r_{core} \ ['']$ \\ \hline
\parbox[0pt][1.5em][c]{0cm}{} galaxy & $762.85^{+40.45}_{-94.09}$ & $32.29^{+0.46}_{-0.33}$ & $30.29^{+0.54}_{-0.26}$ & 0.35 & -49.16 & $1.51^{+0.20}_{-0.19}$ & $0.19^{+0.09}_{-0.86}$ \\ 
\parbox[0pt][1.5em][c]{0cm}{} dark halo & $290.28^{+87.71}_{-66.78}$ & 14.04 & 28.42 & 0.49 & 10.94 & $1.25^{+0.05}_{-0.53}$ & $1.10^{+0.69}_{-0.05}$ \\ \hline
\end{tabular}
\end{center}
\end{table*}

\subsection{Non-parametric -- \pl}
\label{sec:models_pl}

The resolution for the pixel map radius in \pl\ is limited only by computational power and time. The highest, still feasible resolution comprised $\sim 22 \,$ mass pixels corresponding to a physical size of $\sim 3.2'' \times 3.2''$ per pixel.

Furthermore and in analogy to the parametric approach, we did not include all four multiple image systems available in our modelling at once. Instead, we included them again sequentially in order to check for additional predicted images. 

Finally, we included multiple image systems 1 (images 1.1 to 1.5), 2 (images 2.1 to 2.3) and 4 (images 4.1 to 4.3) in full detail. The very complicated eightfold lensed image system 3, however, had to be approximated: the north-western part of image system 3 with the sixfold image (images 3.1 to 3.6) seems to arise due to a complicated caustic folding caused by a local disturbance in the mass distribution, as discussed in Sect. \ref{sec:img_sys3}. Thus, we decided to include only the 'main` images of this north-western part (i.e. images 3.1, 3.4, 3.5 and 3.6) and images 3.7 and 3.8 in our \pl\ analysis.

\begin{figure*}
    \centering
    \begin{minipage}{140mm}
		\centering    	
    	\includegraphics[width=0.5\textwidth]{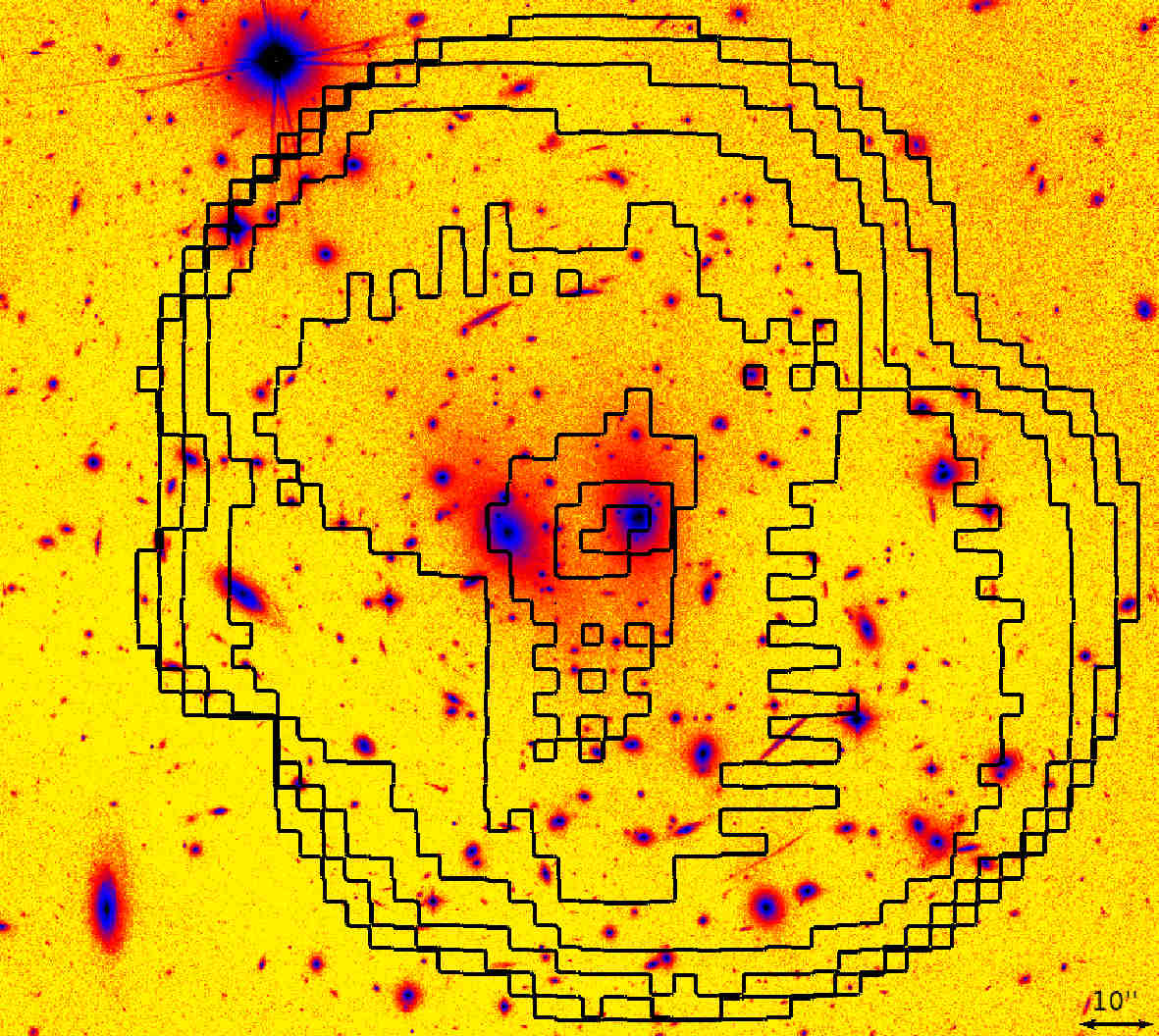}        
    	\caption{Surface mass density contours (solid, black lines) in 
        	     units of the critical surface mass density obtained 
            	 from the best-fitting, non-parametric model in \pl\ for 
             	 the total mass. The spacing between 
             	 contour levels is logarithmic and the difference 
             	 between contour levels is $0.5 \,$mag in surface 
             	 density. The sixth contour from the outside corresponds to the critical density. Contours are overlaid on a composite false 
             	 colour image of RX J1347.5-1145. North is up and East is 
             	 left.}
   		\label{fig:mass_prim_sec_pl}
	\end{minipage}
\end{figure*}

In Fig.~\ref{fig:mass_prim_sec_pl} we show contours of the convergence in logarithmic spacing derived from this mass model. From that already a highly elliptical and irregular mass distribution is visible. 

It is apparent that \pl\ finds significant substructure in the south-eastern part of the cluster. This is coinciding very well with the second NFW component of the \gl\ model and the X-ray observations from \textit{Chandra} (cf. Fig.~\ref{fig:kappa_gl}). As we have noticed in Sect. \ref{sec:intro}, an irregular extension to the south-east of the cluster is visible in X-ray images as well. Thus, this lensing analysis also provides further support for RX J1347.5-1145 being in a merger between two subclusters. 

\subsection{Mass Estimates -- \pl\ \& \gl} 
\label{sec:masses}

From the best-fitting models obtained both with \pl\ and \gl\ we derived estimates for the projected mass enclosed by a cylinder of radius $R$ centred on the BCG:
\begin{equation}
M(<R_n) = \sum_{i=1}^n \kappa (R_i, z) \Sigma_{crit}(z) \pi(R_i-R_{i-1})^2\, \, , 
\end{equation}   
where the convergence $\kappa$ is circularly symmetric and $R_0 = 0$.

\begin{figure*}
    \centering
    \begin{minipage}{140mm}
        \centering
        \includegraphics[width=\textwidth]{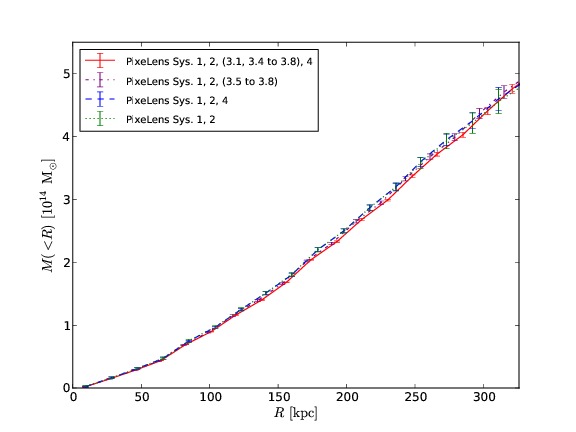}
        \caption{Plots of enclosed mass estimates derived from different 
                 \pl\ models. The models differ by including less and 
                 less images per multiple image system and less and less 
                 multiple image systems in total as indicated in the 
                 legend. All mass estimates are statistically 
                 consistent within their $68 \, \%$ confidence 
                 intervals.}
        \label{fig:different_menc_PL}
    \end{minipage}
\end{figure*}

Apart from deriving a best-fitting model, we also investigated with \pl\ additional mass models derived with different image configurations, i.e. less images per image system and less image systems in total. The results support our simplification regarding image system 3 in the \pl\ analysis (i.e. not including images 3.2 and 3.3; cf. Sect. \ref{sec:models_pl}) since different image configurations do not affect the enclosed mass estimates as shown in Fig.~\ref{fig:different_menc_PL}: all mass estimates are statistically consistent within their $68 \, \%$ confidence intervals. Hence, we conclude that the image systems provide enough constraints for deriving consistent mass estimates with \pl ; as expected, the uncertainties become larger for models with less constraints.
We also want to emphasize that nine out of in total sixteen images are located in a distance range of $150 \, \mathrm{kpc} \leq R_{img} \leq 200 \, \mathrm{kpc}$ from the cluster centre, thus, the most robust estimates for the enclosed mass of the cluster can only be obtained within this range. This argument holds also for the \gl\ analysis.

\begin{figure*}
    \centering
        \centering
        \begin{subfigure}{0.49\textwidth}
            \centering
            \includegraphics[width=\textwidth]{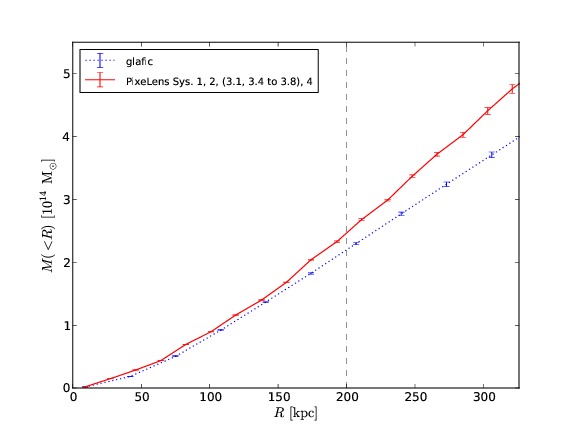}         
            \caption{}
        \end{subfigure}
        \begin{subfigure}{0.49\textwidth}
            \centering
            \includegraphics[width=\textwidth]{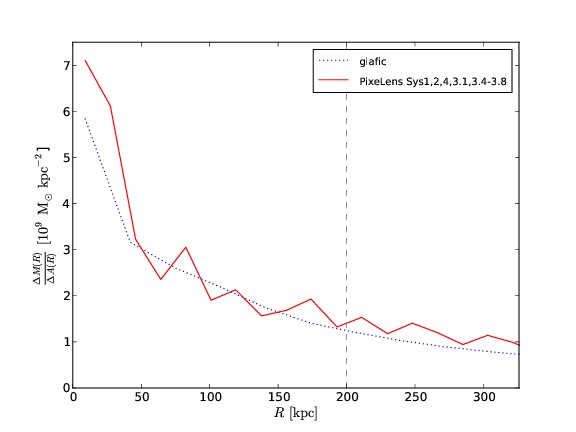}
            \caption{}
        \end{subfigure}
        \caption{\textbf{(a)} Plots of enclosed mass estimates derived from \pl\ 
                 (red, solid line) and \gl\ (blue, dotted line) best-fitting models, 
                 respectively. The dashed, vertical line marks the 
                 radius R = 200 kpc (i.e. distance from cluster centre of multiple image system 2) where deviations in the profiles between \pl\ 	                 and \gl\ arise.
                 \textbf{(b)} Differential mass plots for the best-fitting models of \pl\ (red, solid line) and \gl\ 
                 (dotted, blue line), respectively. The central regions 
                 seem to be systematically overestimated by \pl , and in addition \pl\ assigns also more mass to larger radii. Both combined 				         results in the observed discrepancy between the enclosed mass estimates of \pl\ and \gl , respectively.}
        \label{fig:menc_PL_gl}
\end{figure*}

In Fig.~\ref{fig:menc_PL_gl}a we show the enclosed mass estimates from the best-fitting models of \gl\ and \pl , respectively. The errors for the \gl\ mass estimates were derived using the Monte Carlo approach described in Sect. \ref{sec:errors_gl}. It is apparent that both estimates are consistent within radii $R \lesssim 170 \,$kpc best constrained by multiple image systems, but start to deviate for larger radii where less strong lensing constraints are available. 

Additionally, we show in Fig.~\ref{fig:menc_PL_gl}b the differential mass $\mathrm{d}M/\mathrm{d}R$ plotted against the radius $R$ for the best-fitting models in \pl\ and \gl , respectively. Note, how well the profiles agree overall, just in the central region and for larger radii (less strong lensing constraints) does \pl\ overestimate the mass as compared to \gl .

\begin{figure*} 
    \centering
    \begin{minipage}{140mm}
        \centering
        \includegraphics[width=\textwidth]{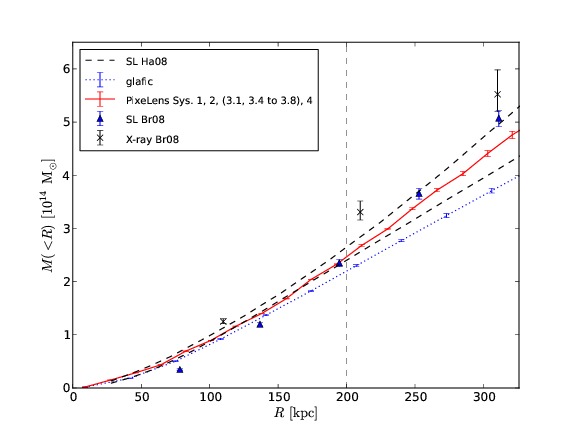}
        \caption{Enclosed mass estimates from strong lensing analyses         
                 using the non-parametric approach of \pl\ (solid, red 
                 line), the parametric approach of \gl\ (blue, dashed 
                 line), a parametrized model (blue, filled triangles) by 
                 \citet{Bradac08} [Br08] and the $68 \, \%$ confidence interval 
                 obtained by \citet{Halkola08} [Ha08] by using strong lensing 
                 data only (black, dashed lines). Additionally, we show a mass 
                 estimate from X-ray data obtained by \citet{Bradac08} (Br08). The dashed, vertical line marks 
                 the position of images 2.2 and 2.3 at a radius $R \sim 200 \, \mathrm{kpc}$ from the cluster centre.}
        \label{fig:mass_all}
    \end{minipage}
\end{figure*}

Finally, we compare the best-fitting mass estimates from \gl\ and \pl\ with the results presented in \citet{Bradac08} and \citet{Halkola08} in Fig.~\ref{fig:mass_all}. Among these results are projected mass estimates from an X-ray data analysis, a parametric strong lensing analysis both obtained by \citet{Bradac08}\footnote{We want to emphasize that we are only comparing to their parametrized strong lensing model and not to their combined (non-parametric) strong and weak lensing model, since our focus is on the cluster core only.} and the $68 \, \%$ confidence interval based on strong lensing data from \citet{Halkola08}. While the estimates from strong lensing analyses by \citet{Bradac08} and by \citet{Halkola08} and from \pl\ agree well within their uncertainties, the estimates from \gl\ appear to be significantly lower, especially for radii $R \gtrsim 150 \, \mathrm{kpc}$. The estimates based on X-ray measurements performed by \citet{Bradac08} are somewhat high compared to the strong lensing analyses.

In order to give a quantitative example, we use the distance $R_{arc} \sim 35'' \sim 200 \, \mathrm{kpc}$ from the prominent, elongated arc (images 2.2 and 2.3 at redshift $z=1.75$) to the cluster centre for an estimate of the enclosed, projected mass $M(<R_{arc})$ within that radius. The best fit from \gl\ yields a mass $M(<R_{arc}) \approx (2.19^{+0.01}_{-0.02}) \times 10^{14} \, \mathrm{\Msol}$, whereas the \pl\ estimate exceeds this by $\sim 13 \, \%$ with a mass of $M(<R_{arc}) \approx (2.47 \pm 0.01) \times 10^{14} \, \mathrm{\Msol}$. The \pl\ estimate is statistically consistent with the estimates by \citet{Halkola08} $M(<R_{arc}) \approx (2.56 \pm 0.12) \times 10^{14} \, \mathrm{\Msol}$ and \citet{Bradac08} $M(<R_{arc}) \approx (2.47^{+0.06}_{-0.07}) \times 10^{14} \, \mathrm{\Msol}$. However, the X-ray mass estimate $M(<R_{arc}) \approx (3.10^{+0.19}_{-0.14}) \times 10^{14} \, \mathrm{\Msol}$ within this radius as estimated by \citet{Bradac08} is higher by $\sim 26 \, \%$ compared to the values from \pl\ and \citet{Halkola08} and it exceeds the result from \gl\ by $\sim 42 \, \%$. Note, however, that the strong lensing analyses by \citet{Bradac08} and \citet{Halkola08} make different assumptions about the lensing data (cf. Tab.~\ref{tab:arcs}).

\section{Conclusion}
\label{sec:conclusion}

Based on image identifications and strong lensing analyses in \citet{Bradac08} and \citet{Halkola08}, as well as by interpreting some of the multiple image systems anew and including new images in the analysis, we present a consistent strong lensing analysis of the cluster RX J1347.5-1145.

We have reconstructed its mass distribution by employing the parametric software \gl\ \citep{Oguri10} and the non-parametric software \pl\ \citep{Saha97, Saha04}.

The results from these two analyses present further support for the merger scenario and produced mass maps of the cluster which agree well with each other in revealing several mass components and a highly elliptical mass distribution. Furthermore, the fitted position of the second NFW profile from our parametric \gl\ model coincides very well with a region of shocked gas visible in X-ray data \citep{Komatsu01}. Also the non-parametric best-fitting model obtained with \pl\ assigns a high amount of substructure to this cluster region. We find that the mass estimates of \pl\ and \gl\ are consistent within $\sim 13\% $ at radii best constrained by our data, but deviate stronger for larger radii with less constraints. 

This difference inside the arc radii could be due to on the one hand, that the parametric model used in \gl\ could be inaccurate in the sense of not assigning sufficient mass to the profiles in use in this model (or just not containing sufficient additional profiles). This would imply that this model is not finding physically existent mass in the outer regions of the cluster which is in contrast well-captured by the other analyses (cf. Fig~\ref{fig:mass_all}).

On the other hand, the mass model in \gl\ is the only one among all these analyses that explains the images 3.1 to 3.8 as resulting from only one source. Considering the consistent redshifts of these images according to CLASH data supports this approach strongly. The eightfold occurrence of the same image is consistent with being caused by a complicated swallowtail caustic folding. Such a complicated folding can be caused by two additional mass components in the vicinity of images 3.1 to 3.6. One of these represents a mass overdensity in the north-western part of the cluster which is coinciding with a higher concentration of cluster galaxies in this region. The other perturbing profile is less massive and might be physically connected to a very faint background object. Hence, introducing these additional components which leads to the corresponding caustic folding, might be the reason for this particular mass model to require less mass than the other models.

Note, however, that \citet{SchneiderSluse13} have shown that strong lensing observables (except time delays) are invariant under the so-called ''source-plane transformation`` which leads to a certain arbitrariness in the choice of mass models. This invariance is only an approximate one for asymmetric lenses such as clusters and the comparison of our findings to previous results obtained with less well-constrained data (e.g. redshifts) and different assumptions about the mass profile shows that all these mass profiles are consistent albeit within a larger scatter not accounted for by the estimated uncertainties (cf. Fig~\ref{fig:mass_all}). This degeneracy in mass profiles might thus be another hint towards such a fundamental level of systematic uncertainties immanent in strong lensing due to the source-plane transformation. 

To measure cluster masses beyond the arc regime, X-ray masses or weak lensing estimates are required. In this respect it is also important to note that the weak lensing studies at larger radii by \citet{FischerTyson97} and the one included in \citet{Bradac08} are consistent with X-ray measurements by \citet{Allen02} and \citet{Bradac08}, respectively.

\section*{Acknowledgments}

We would like to thank the anonymous referee for detailed comments which helped to further improve this work and its presentation.

This work profits immensely from the high-quality data made available by CLASH. We thank the whole CLASH team for their amazing effort and their community-friendly data release policy.
We are deeply grateful to P. Saha and M. Oguri for making their software available to the public and especially for their helpful suggestions and correspondence.  
We also thank J. Wambsgan\ss\ and H. Hoekstra for fruitful comments on this work.

\bibliographystyle{mn2e} 
\bibliography{bibliography}
\label{lastpage}

\end{document}